\documentclass[aps,prx,twocolumn,superscriptaddress,nofootinbib]{revtex4-2}
\usepackage{mathrsfs}
\usepackage{amsmath,amssymb,bm,mathtools}
\usepackage{blindtext}
\usepackage{graphicx}
\graphicspath{ {figures/}{figures/eps/}{figures/pdf/} }
\usepackage{array}
\usepackage{float}

\usepackage[export]{adjustbox}
\usepackage{framed}

\DeclareUnicodeCharacter{2212}{-}

\begin{document}
\title{Magnetic properties of the quasi-XY Shastry-Sutherland magnet Er$_2$Be$_2$SiO$_7$}

\author{A. Brassington} 
\affiliation{Department of Physics and Astronomy, University of Tennessee, Knoxville, TN 37996, USA}

\author{Q. Ma} 
\affiliation{Neutron Scattering Division, Oak Ridge National Laboratory, Oak Ridge, TN 37831, USA}

\author{G. Sala} 
\affiliation{Oak Ridge National Laboratory, Oak Ridge, TN 37831, USA}

\author{A.I. Kolesnikov} 
\affiliation{Neutron Scattering Division, Oak Ridge National Laboratory, Oak Ridge, TN 37831, USA}

\author{K.M. Taddei} 
\affiliation{Neutron Scattering Division, Oak Ridge National Laboratory, Oak Ridge, TN 37831, USA}

\author{Y. Wu} 
\affiliation {Neutron Scattering Division, Oak Ridge National Laboratory, Oak Ridge, TN 37831, USA}

\author{E.S Choi} 
\affiliation {National High Magnetic Field Laboratory and Department of Physics, Florida State University, Tallahassee, Florida 32310, USA}

\author{H. Wang} 
\affiliation {Department of Chemistry, Michigan State
University, East Lansing, Michigan 48824, United States}

\author{W. Xie} 
\affiliation {Department of Chemistry, Michigan State
University, East Lansing, Michigan 48824, United States}

\author{J. Ma} 
\affiliation {Key Laboratory of Artificial Structures and Quantum Control, Shenyang National Laboratory for Materials Science, Shanghai Jiao Tong University, Shanghai 200240, China}

\author{H.D. Zhou} \altaffiliation{\href{mailto:hzhou10@utk.edu}{hzhou10@utk.edu}}
\affiliation{Department of Physics and Astronomy, University of Tennessee, Knoxville, TN 37996, USA}

\author{A.A. Aczel} \altaffiliation{\href{mailto:aczelaa@ornl.gov}{aczelaa@ornl.gov}}
\affiliation{Neutron Scattering Division, Oak Ridge National Laboratory, Oak Ridge, TN 37831, USA}
 
\date{\today}

\begin{abstract}

Polycrystalline and single crystal samples of the insulating Shastry-Sutherland compound Er$_2$Be$_2$SiO$_7$ were synthesized via a solid-state reaction and the floating zone method respectively. The crystal structure, Er single ion anisotropy, zero-field magnetic ground state, and magnetic phase diagrams along high-symmetry crystallographic directions were investigated by bulk measurement techniques, x-ray and neutron diffraction, and neutron spectroscopy. We establish that Er$_2$Be$_2$SiO$_7$ crystallizes in a tetragonal space group with planes of orthogonal Er dimers and a strong preference for the Er moments to lie in the local plane perpendicular to each dimer bond. We also find that this system has a non-collinear ordered ground state in zero field with a transition temperature of 0.841 K consisting of antiferromagnetic dimers and in-plane moments. Finally, we mapped out the $H-T$ phase diagrams for Er$_2$Be$_2$SiO$_7$ along the directions $H \parallel$ [001], [100], and [110]. While an increasing in-plane field simply induces a phase transition to a field-polarized phase, we identify three metamagnetic transitions before the field-polarized phase is established in the $H \parallel$ [001] case. This complex behavior establishes insulating Er$_2$Be$_2$SiO$_7$ and other isostructural family members as promising candidates for uncovering exotic magnetic properties and phenomena that can be readily compared to theoretical predictions of the exactly soluble Shastry-Sutherland model. 

\end{abstract}  

\maketitle 

\section{Introduction}

Magnetic systems with a combination of geometric frustration and antiferromagnetic exchange interactions have attracted significant attention as candidates for hosting novel magnetic states with intrinsically quantum mechanical properties \cite{94_ramirez, 00_greedan, 06_moessner, 14_mirebeau}. In such frustrated magnets competing exchange interactions prevent the system from finding a global arrangement of spins which minimizes energy. This suppression of long-range magnetic order allows for the formation of novel phases such as quantum spin liquids (QSLs) \cite{10_balents, 16_norman, 17_savary, 17_zhou, 19_knolle, 19_takagi, 19_wen, 20_broholm, 21_chamorro} characterized by long-range entanglement and capable of hosting exotic quasi-particle excitations. Frustrated magnets tend to have complicated magnetic phase diagrams with a variety of different states stabilized for various combinations of applied magnetic field and temperature. Notable examples of frustrated geometries include the kagome \cite{22_yin}, pyrochlore \cite{06_greedan, 10_gardner}, and triangular lattices \cite{15_starykh}. 

The geometrically-frustrated, quasi-two-dimensional Shastry-Sutherland lattice (SSL) is unique due to its realization of a magnetic Hamiltonian that is exactly solvable \cite{Shastry_1981}. One can generate the SSL by starting with a square lattice and then adding an additional diagonal bond to each ion. This results in  structure where each site has one nearest-neighbor (NN) with coupling strength $J$ and four next-nearest-neighbors each with coupling strength $J'$. Critically, the magnetic ground state of SSL systems depends both on the effective spin of the magnetic ion and the ratio $J'/J$ \cite{Shastry_1981}. For the effective spin-1/2 case with antiferromagnetic exchange interactions, when $J'/J$ is small the ground state is known to be a product of dimer singlets. As $J'/J$ increases the ground state of the system transitions through at least one and perhaps several intermediate phases, possibly including the long sought after QSL. Finally, when $J'/J$ is large the ground state is the same antiferromagnetic Neel spin configuration expected for a simple square lattice.

The small $J'/J$ limit is exemplified by the well-studied insulating compound SrCu$_2$(BO$_3$)$_2$ whose low-temperature ground state consists of a series of spin-singlet dimers \cite{Miyahara_1999, Kageyama_1999}. Furthermore, SrCu$_2$(BO$_3$)$_2$ is close to the critical $J'/J$ value and it can be driven across the quantum phase transition into an antiferromagnetic Neel state by the application of pressure \cite{Guo_2020, Sakurai_2009, Haravifard_2016}. At pressures between these two extremes an intermediate singlet plaquette phase whose basic unit consists of four entangled spins is observed and multiple theoretical studies predict the existence of another intermediate QSL phase \cite{Yang_2022, Wang_2022}. Under extreme magnetic fields SrCu$_2$(BO$_3$)$_2$ displays several magnetization plateaus at rational fractions of the saturation magnetization, which are attributed to the crystallization of the excited triplets \cite{Miyahara_2000, Kageyama_2000, Miyahara_2000_1, 08_sebastian}. Overall, SrCu$_2$(BO$_3$)$_2$ provides a rich platform for exploring several novel and exotic magnetic phenomena. 

Since SrCu$_2$(BO$_3$)$_2$ is a particularly fascinating system and theoretical work on the SSL abounds, there is significant interest in identifying and characterizing other magnetic materials that realize the $J'/J$ $S =$~1/2 SSL model. Rare-earth-based materials are particularly attractive because crystal field effects often generate $J_{\rm eff} =$~1/2 moments with a variety of magnetic anisotropies. The $R$B$_4$ ($R =$~rare earth) family was the first group of materials studied in this context \cite{03_ji, 05_watanuki, 05_wigger, 06_yoshii, 06_michimura, 07_matsumura, 07_okuyama, 10_kim, 17_yamauchi}. However for $R$B$_4$ the crystal field schemes are not well-known, the zero-field magnetic ground states are typically ordered \cite{81_fisk, 81_gianduzzo, 81_will, 05_wigger, 06_michimura, 07_okuyama, 07_iga, 08_okuyama, 16_sim, 17_yamauchi, 06_blanco}, and their metallic behavior ensures that the simple $J'/J$ SSL model is not appropriate due to the extended nature of the RKKY exchange interactions. Similar issues were identified in subsequent metallic SSL systems with the general chemical formula $R_2T_2X$ ($T = $~Pd, Ge, Ni, Cu, Pt, Si, Rh, and $X =$~Mg, In, Cd, Sn, Pb, Al) \cite{95_hulliger, 98_giovannini, 03_kraft, 04_zaremba, 06_rayapol, 08_kumar, 09_schappacher, 09_shah, 11_suen, 12_shimura, 16_miiller, 16_wu, 18_gannon, 19_gannon}.

More recently, insulating SSL systems based on rare-earth ions have been discovered. One such family is Ba$R_2 M_2$X$_5$ ($M =$~Zn, Pd, Pt; $X =$~O, S) \cite{03_wakeshima, 08_ozawa, 20_ishii, 21_ishii, 22_billingsley, 23_marshall, 23_pasco} and a second family is $R_2$Be$_2$GeO$_7$ \cite{21_ashtar} or $R_2$Be$_2$SiO$_7$ \cite{Brassington_2024}. Limited characterization work has been reported on Ba$R_2 M_2$X$_5$, with most studies concentrating on the Nd-based systems and identifying $J_{\rm eff} =$~1/2 moments, ferromagnetic dimers, and zero-field ordered ground states with predominantly in-plane spin configurations \cite{21_ishii, 22_billingsley, 23_marshall}. On the other hand, nothing beyond the initial discovery paper has been reported for $R_2$Be$_2$GeO$_7$ or $R_2$Be$_2$SiO$_7$, with the exception of some recent bulk characterization work on Nd$_2$Be$_2$GeO$_7$ and Pr$_2$Be$_2$GeO$_7$ \cite{24_liu}.

Here, we work towards addressing this issue by investigating the magnetic properties of polycrystalline and single crystal samples of Er$_2$Be$_2$SiO$_7$ with a combination of dc and ac magnetic susceptibility, magnetization, heat capacity, x-ray and neutron diffraction, and neutron spectroscopy measurements. We find that the Er moments exhibit quasi-XY magnetic anisotropy with a strong tendency to lie in the local plane perpendicular to each dimer bond. The system hosts non-collinear magnetic order with in-plane moments and antiferromagnetic dimers below 0.841~K. While an in-plane magnetic field simply polarizes the moments, an $H \parallel [001]$ field induces three metamagnetic transitions before the field-polarized phase is realized. The insulating behavior of Er$_2$Be$_2$SiO$_7$ ensures that its complex magnetic properties measured in the laboratory can be compared to theoretical predictions of the $J'/J$ SSL model.  

\begin{figure*}
\scalebox{0.4}{\includegraphics{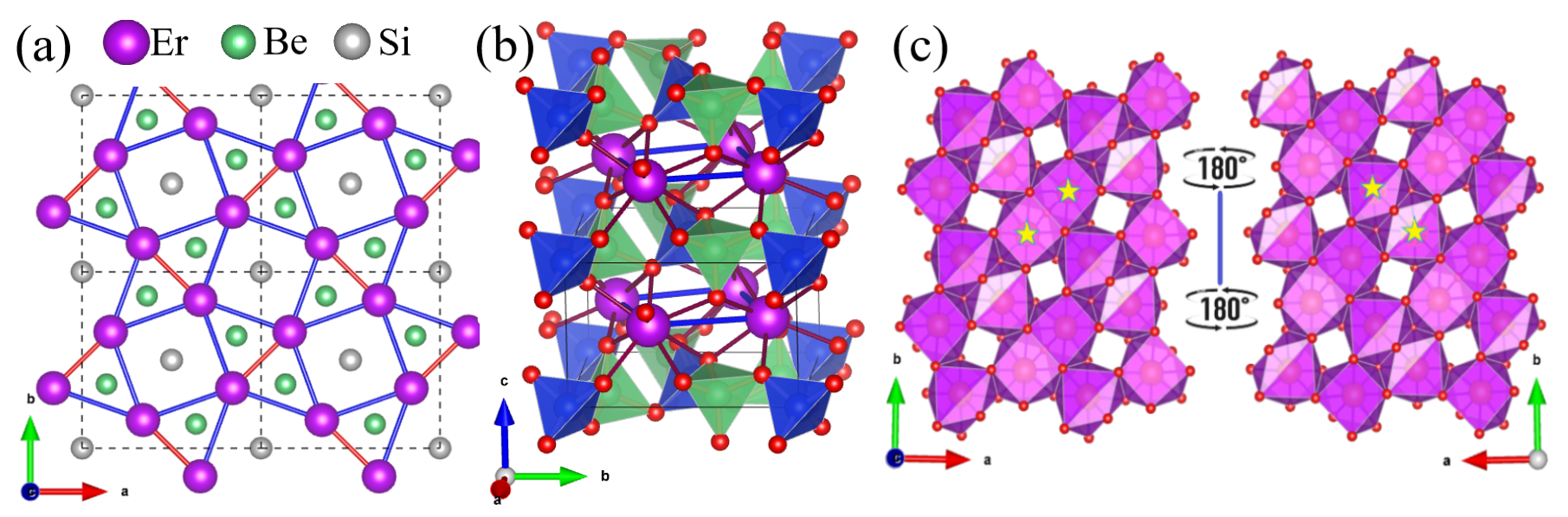}}
\caption{(a) Schematic diagram of the Er$^{3+}$ ions forming a SSL lattice in the $ab$-plane. Intradimer bonds $J$ (3.265~\AA) are shown in red and interdimer bonds $J'$ (3.862~\AA) in blue . The relative positions of the Be and Si ions are also shown. (b) Structure of Er$_2$Be$_2$SiO$_7$ viewed slightly off the [100] axis. The SSL layers formed by the ErO$_8$ polyhedra are separated by layers of BeO$_4$ (green) and SiO$_4$ (blue) tetrahedra. (c) The monoclinic point symmetry ($C_s$) of the Er ions is best illustrated by viewing the crystal structure along the [001] and [00-1] directions. A single NN pair of Er ions is marked with stars for reference. The normal to the mirror plane lies along the crystallographic [110] or [1-10] direction for the orthogonal dimer sublattices in the system.}
\label{Fig1}
\end{figure*}

\section{Experimental Methods}
Er$_2$Be$_2$SiO$_7$ single crystals were grown using the floating zone method from a packed, phase pure polycrystalline sample with a growth speed of 10mm/h \cite{Brassington_2024}. To synthesize the polycrystalline sample, stoichiometric amounts of Er$_2$O$_3$ (Strem Chemicals, $99.999\%$), SiO$_2$ (Alfa Aesar, $99.8\%$), and BeO (Alfa Aesar, $99\%$) were ground and packed into a long cylindrical rod. This rod was first annealed in air at $1000^{\circ}$C and then twice more at $1400^{\circ}$C for 20 hrs each time with intermediate grindings. A similar procedure was used to synthesize polycrystalline Lu$_2$Be$_2$SiO$_7$ with Lu$_2$O$_3$ (Alfa Aesar, 99.99$\%$) substituted for Er$_2$O$_3$.

The phase purity of both the powders and single crystals was confirmed by performing room temperature powder x-ray diffraction (XRD) using a HUBER imaging plate Guinier camera 670 with Cu radiation ($\lambda = $~1.54059~\AA). The single crystal samples were first crushed for these measurements. Refinement of the XRD pattern was done using the software package FULLPROF \cite{93_rodriguez} using the structure of Y$_2$Be$_2$SiO$_7$ \cite{Kuz'micheva2002} as a starting reference. Single crystal x-ray diffraction (SXRD) measurements were carried out on a Bruker Eco Quest X-ray Diffractometer using Mo radiation ($\lambda =$~0.71073~\AA) at 300 K and the refinement was done using the Bruker SHELXTL Software Package. The main results are presented in Table~\ref{table:SXRD2} and the results agree well the previous powder refinement \cite{Brassington_2024}. For all directional-dependent measurements, the crystal orientation was determined using the Laue backscattering method. 

Temperature and field-dependent magnetization data were collected at the University of Tennessee using a Materials Property Measurement System (MPMS) from Quantum Design with a vibrating squid magnetometer (VSM) option. Magnetization measurements at He-3 temperatures were carried out in the same MPMS using a He-3 insert also from Quantum Design. Heat capacity measurements under field were carried out using a Physical Property Measurement System (PPMS) Dynacool with a He-3 insert at Shenyang National Laboratory. AC susceptibility measurements were performed at the National High Magnetic Field Laboratory (NHMFL) with both He-3 and dilution refrigerator configurations. 

Neutron powder diffraction was performed using the high-resolution HB-2A powder diffractometer \cite{18_calder} at the High Flux Isotope Reactor (HFIR) of Oak Ridge National Laboratory (ORNL). The $\sim$~4~g powder sample was sealed inside an aluminum can with 1~atm of helium exchange gas. Diffraction patterns were measured at 280 mK and 4 K with a wavelength of 2.41~\AA~and a collimation of open-open-12'. Inelastic neutron scattering on $\sim$~3~g of similarly prepared powder was performed on the direct geometry time-of-flight chopper spectrometer SEQUOIA \cite{10_granroth} at the Spallation Neutron Source (SNS) at ORNL. Data were collected at 5 and 25 K with incident energies of 25 and 80 meV. For these measurements, the fine Fermi chopper was spun at 240 or 480~Hz and the $T_0$ chopper setting was 60 or 90~Hz. Spectra were collected with the same instrument settings on a similar mass of the non-magnetic reference sample Lu$_2$Be$_2$SiO$_7$ to facilitate a background subtraction of the phonon spectra. 
 
A single crystal neutron diffraction experiment was performed on the DEMAND (HB-3A) beamline \cite{19_cao} at the HFIR with an incident wavelength of 1.542~\AA. A single crystal sapmle with a mass $\sim$~30 mg was measured in two-axis mode in a vertical-field cryomagnet down to temperatures of 280 mK using a He-3 insert. A magnetic field of up to 2 T was applied along the [001]-axis.

\section{Results and Discussion}

\subsection{Crystal Structure}

Er$_2$Be$_2$SiO$_7$, like all members of the R$_2$Be$_2$SiO$_7$ family, crystallizes into a tetragonal lattice with space group $P-42_1m$ (113). The room-temperature lattice constants are $a =$~7.2572(4)~\AA~and $c =$~4.7396(3)~\AA. The structure consists of planes of eight-fold coordinated Er$^{3+}$ ions with spacer layers containing both Si-O and Be-O tetrahedra. The sublattices formed by Er, Be, and Si all stack in an AAA-type fashion along the [001]-axis. The magnetic Er$^{3+}$ and non-magnetic Be$^{2+}$ ions both form SSLs in the $ab$-plane, while the non-magnetic Si$^{4+}$ ions form a square lattice. The NN and NNN in-plane distances at room temperature are 3.265~\AA~and 3.862~\AA~respectively while the shortest interplane distance is 4.740~\AA, which implies a quasi-two-dimensional magnetic system. Schematics of the crystal structure viewed along the [001]-axis and just slightly off the a-axis are shown in Figs.~\ref{Fig1}(a) and (b), with an outline of the chemical unit cell also shown in the latter. 


\begin{table}
\caption{ {\footnotesize Atomic coordinates and isotropic displacement parameters $U_{eq}$ (\AA$^2$). All sites are fully occupied.}}

\begin{tabular}{c c c c c  c} 
\hline \hline
 Atom   & Wyck. & x \hspace{1 pt} & \hspace{1 pt} y \hspace{1 pt} & \hspace{1 pt} z \hspace{1 pt} & \hspace{1 pt}$U_{eq}$ \hspace{1 pt}\\
\hline
Er & 4e & 0.15907(3) & 0.65907(3) & 0.50719(8) & 0.0041(19)\\
\hline
Si & 2a & 0 & 0 & 0  & 0.0036(5)\\
\hline
Be & 4e &
0.6352(14) &0.1352(14) &0.960(3)  & 0.0060(18)\\
\hline
O$_1$ & 2c & 0 & 1/2 & 0.180(2) & 0.0069(17)\\
\hline
O$_2$ &4e & 0.6411(7) & 0.1411(7) &0.2934(5)  & 0.0041(11)\\
\hline
O$_3$ & 8f & 0.0816(8) & 0.1634(7) & 0.2017(10)  & 0.0050(8)\\
\hline \hline

\end{tabular}
\label{table:SXRD2} 
\end{table}

\begin{figure}
\centering
\scalebox{0.43}{\includegraphics{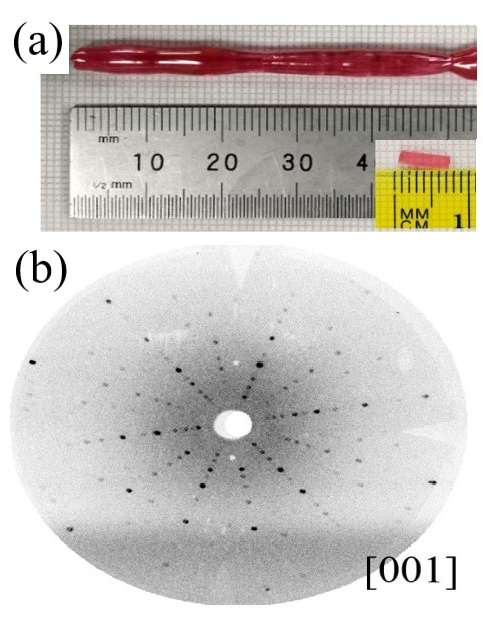}}
\caption{(a) An as-grown crystal boule of Er$_2$Be$_2$SiO$_7$. (b) The Laue pattern generated with the incident x-ray beam along the crystallographic [001]-axis.}
\label{Fig2}
\end{figure}

The Er and Be ions occupy two different 4e Wyckoff sites, while Si occupies the 2a Wyckoff site. The ligand environment around the Er$^{3+}$ ions are irregularly-shaped polyhedra consisting of eight O$^{2-}$ ions, as shown in Fig.~\ref{Fig1}(c). Opposite faces of the polyhedra, located above and below the SSL planes, have distorted square and diamond geometries respectively. The Er$^{3+}$ ions have a monoclinic point group symmetry of C$_s$, which means that their only symmetry element is a mirror plane. The Er ions forming the two different types of dimers have different directions for their mirror plane normals corresponding to the crystallographic [110] and [1-10] axes.  The ErO$_8$ polyhedron shares a face with its one NN polyhedron and an edge with each of its four NNN polyhedra. Detailed crystal structure refinement results obtained from single crystal XRD are given in Table~\ref{table:SXRD2}. An image of a floating zone crystal boule and a Laue pattern with the beam oriented along the crystallographic [001]-axis are shown in Fig.~\ref{Fig2}. 

\subsection{Single Ion Anisotropy and Crystal Fields}

\begin{figure*}
\centering
\scalebox{1}{\includegraphics{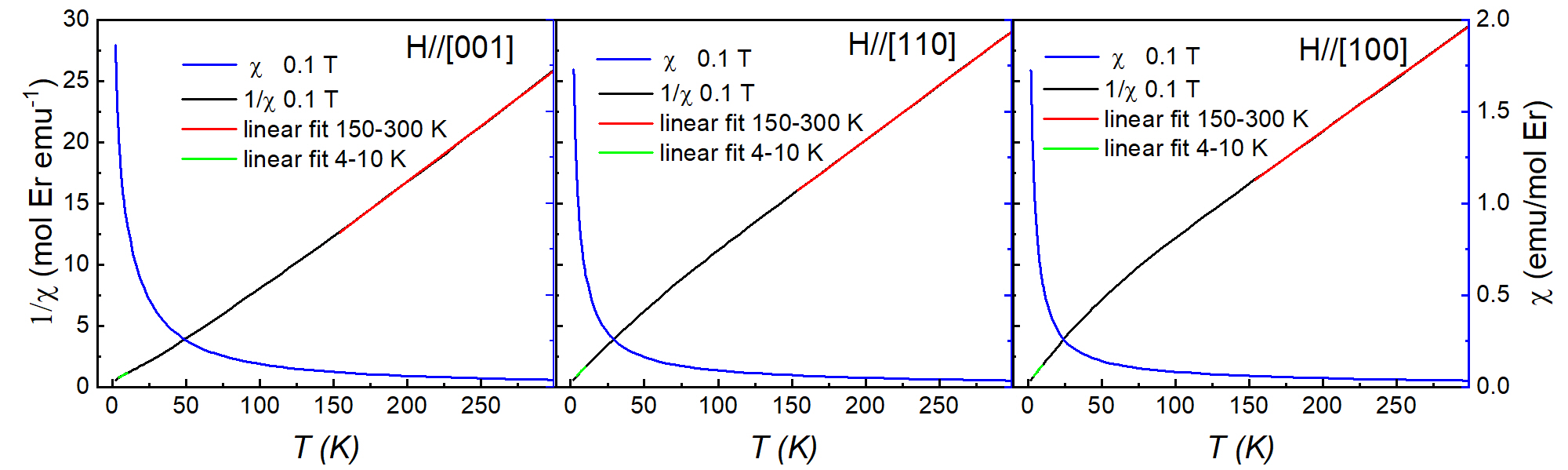}}
\caption{$1/\chi$ (left) and $\chi$ (right) vs $T$ for $H \parallel [001]$, $H \parallel [110]$, and $H \parallel [100]$ collected at 0.1~T. Linear fits are performed between 150-300~K and 4-10~K. The obtained $\theta_{cw}$ and $\mu_{\rm eff}$ values are noted in the text.}
\label{Fig3}
\end{figure*}

Figure~\ref{Fig3} shows the $T$-dependence of both the magnetic susceptibility $\chi$ (plotted as $M/H$) and its inverse 1/$\chi$ along three high-symmetry crystallographic directions. 1/$\chi$ does not have a linear $T$-dependence throughout the entire range measured due to the increased depopulation of excited Er$^{3+}$ crystal field levels with decreasing temperature. The 1/$\chi$ data was first fit to a Curie-Weiss law in the high-temperature range 150-300~K and we found that the Er$^{3+}$ effective moment agrees well with expectations for a free ion system, with a small variation for the different datasets. The values were 9.3597(4)~$\mu_B$ for $H \parallel$~[001], 9.4618(2)~$\mu_B$ for $H \parallel$~[110], and 9.6158(2)~$\mu_B$ for $H \parallel$~[100]. We proceeded to perform another fit to the same Curie-Weiss function in the low-temperature range 4-10~K to get a rough estimate of the Curie-Weiss temperature $\theta_{CW}$. We found that it was negative for all three datasets, with values of -6.60(4)~K for $H \parallel$~[001], -2.66(3)~K for $H \parallel$~[110], and -1.74(3)~K for $H \parallel$~[100]. The sign of $\theta_{CW}$ indicates that Er$_2$Be$_2$SiO$_7$ has dominant antiferromagnetic exchange interactions, which is likely due to an antiferromagnetic intradimer coupling.

\begin{figure}
\centering
\scalebox{0.60}{\includegraphics{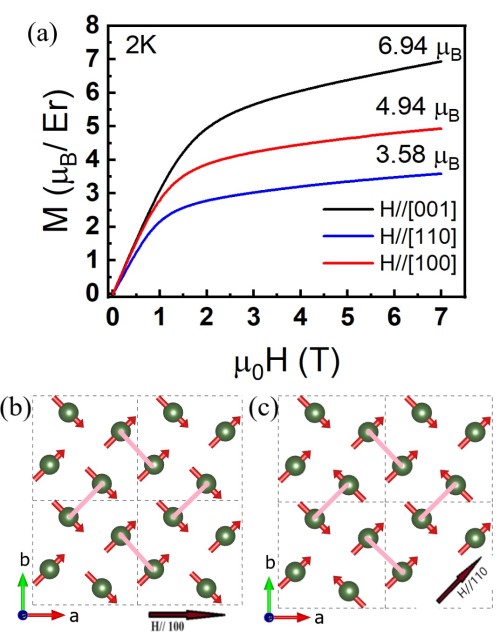}}
\caption{(a) DC magnetization ($M$) vs field ($H$) for the three high-symmetry directions $H \parallel [001]$, $H \parallel [100]$, and $H \parallel [110]$. The magnetization values at 7 T are provided in the panel. (b) Predicted field-polarized spin structure for $H \parallel [100]$. The moments are confined to the $ab$-plane and make a 45$^\circ$ angle with the local [100] axis. (c) Predicted field-polarized spin structure for $H \parallel [110]$. Due to in-plane anisotropy with a strong tendency for the moments to point along local [110] directions, only one sublattice is polarized while the other one is essentially unresponsive.}
\label{Fig4}
\end{figure}

Figure~\ref{Fig4}(a) shows DC magnetization vs applied field at 2~K. Interestingly, the saturation magnetization ratio between the [110] and [100] directions, $M_{sat}^{110}/M_{sat}^{100}$, is almost exactly sin(45$^{\circ}$) $=$ 1/$\sqrt{2}$. This observation can be used to draw some general conclusions about the Er$^{3+}$ single ion anisotropy in this system. Since the SSL can be described as two interpenetrating square networks of orthogonal dimers, the measured anisotropy in the $ab$-plane magnetization can arise from a combination of a [110] easy axis and a [1-10] hard axis for one dimer sublattice and vice versa for the second dimer sublattice. A [110] applied field would then only polarize one sublattice while the other would not respond to it at all. For a [100] applied field, both sublattices would contribute to the field-polarized state in an equivalent manner, as all moments would have a component along the field direction but they would remain parallel to their [110] easy-axis direction. Schematics of the expected field-polarized states for the [100] and [110] applied field directions that are consistent with the anisotropic saturation magnetization values are shown in Fig.~\ref{Fig4}(b) and (c). Similar in-plane anisotropy has been observed in other magnetic systems with a SSL geometry, including Yb$_2$Pt$_2$Pb \cite{12_shimura} and BaNd$_2$ZnS$_5$ \cite{22_billingsley}. 

The saturation magnetization ratio of $M_{\rm sat}^{110}/M_{\rm sat}^{001} \approx$~1/2 provides additional insight into the Er$^{3+}$ single ion anisotropy. Based on the established $ab$-plane anisotropy, the [001] saturation magnetization can be explained by a simple field-polarized state with the moments from both sublattices aligned with the field. Since the low-field magnetization slopes are nearly equivalent for the [110] and [001] field directions when scaling the former dataset to account for the decreased number of moments contributing to the response, Er$_2$Be$_2$SiO$_7$ appears to be an XY (or quasi-XY) magnet. This behavior differs strongly from observations for Yb$_2$Pt$_2$Pb \cite{Ochiai_2011, Gannon_2019} and BaNd$_2$ZnS$_5$ \cite{22_billingsley}, where the $c$-axis turned out to be a hard direction for the magnetization. 

To investigate quantitative details of the magnetic anisotropy induced by crystal fields, inelastic neutron scattering (INS) measurements on a polycrystalline sample of Er$_2$Be$_2$SiO$_7$ were conducted using the time-of-flight chopper spectrometer SEQUOIA. Datasets were collected with incident energies of 25 meV and 80 meV at temperatures of 5 and 25 K to measure the crystal field levels associated with the $J =$~15/2 ground state multiplet of Er$^{3+}$ and the virtual transitions created by thermally-populating the lowest-lying crystal field level. Er$^{3+}$ is a Kramers ion, so the crystal field ground state must be at least two-fold degenerate. As shown in Figs.~\ref{Fig1}(b) and (c), each Er ion is surrounded by eight oxygen ions that generate a monoclinic $C_s$ environment. The low-symmetry of this point group is expected to produce maximal splitting of the 16 levels associated with the $J =$~15/2 ground state multiplet, so up to 7 excited crystal field doublets may be visible in the low-temperature INS data. 

\begin{figure}
\centering
\scalebox{0.39}{\includegraphics{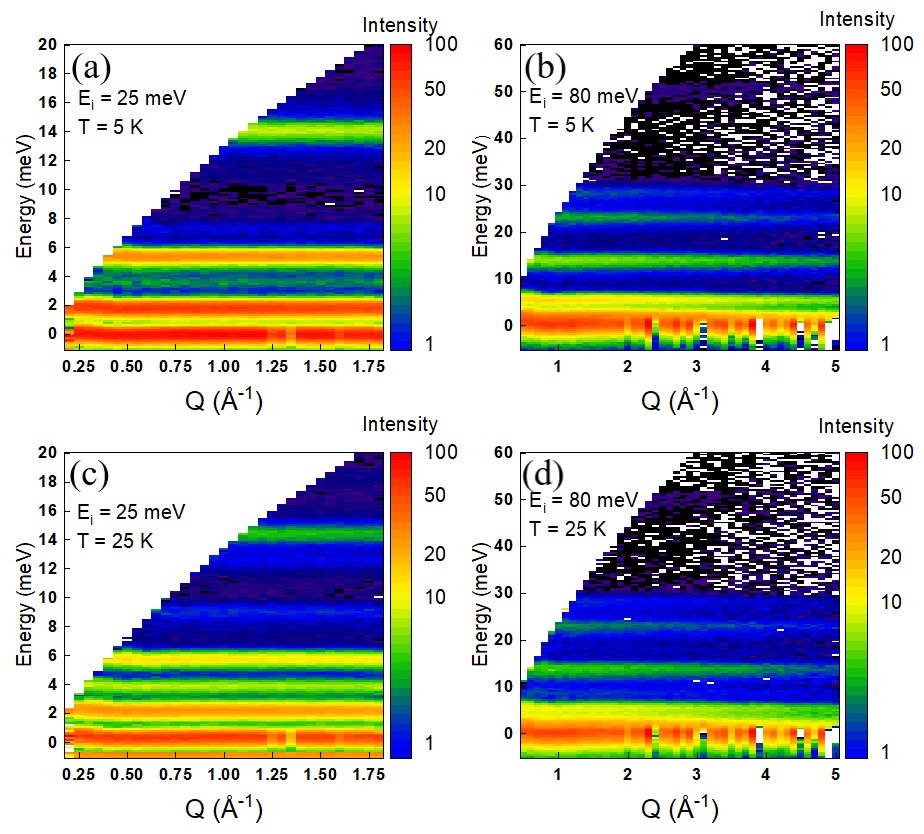}}
\label{SEQ_Spec}
\caption{Color contour plots of the scattering intensity (in arb. units) as a function of momentum transfer $Q$ and energy transfer $E$ for the following datasets: (a) $E_i =$~25~meV, $T =$~5 K, (b) $E_i =$~80~meV, $T =$~5 K, (c) $E_i =$~25~meV, $T =$~25 K, (d) $E_i =$~80~meV, $T =$~25 K. Several crystal field excitations from Er$^{3+}$ are visible in this data. }  
\label{Fig5}
\end{figure}
 
Figures~\ref{Fig5}(a) and (b) present color contour plots of the $E_i =$~25~meV and 80~meV datasets respectively at 5 K. The phonon spectra from a non-magnetic reference sample of Lu$_2$Be$_2$SiO$_7$ were subtracted off from these data. There are three strong, flat modes in the lower incident energy dataset centered about energy transfers of 1.77 meV, 5.25 meV, and 13.72 meV, with two additional strong modes visible in the other dataset centered about energy transfers of 22.53 meV and 27.62 meV. Three weaker modes are also present at energy transfers of 3.48 meV, 7.17 meV, and 49.32 meV. All eight modes have a magnetic origin, as their intensities decrease monotonically with increasing $Q$. The higher temperature data collected at 25 K is shown in Figs.~\ref{Fig5}(c) and (d) and proved invaluable for elucidating the crystal field scheme. While seven of the eight modes previously observed had reduced intensity at 25 K, the intensity of the 3.48 meV mode was enhanced. This finding suggests that this mode corresponds to a virtual crystal field transition. Indeed, the energy of this mode corresponds to the difference between the first and second excited doublets. Therefore, the other seven levels observed at 5 K are the expected transitions associated with the ground state doublet. There are several other virtual crystal field transitions observed in the 25 K datasets as well, with the most prominent features centered at 8.29 meV and 11.81 meV. The energy levels and integrated intensities of these CEF excitations have been extracted by fitting constant $Q$-cuts (integration range of 0.2 to 3~\AA$^{-1}$) to a sum of Lorentzian functions with a linear background. The integrated intensities of the crystal field modes above 20 meV were normalized by a scale factor obtained by comparing the integrated intensities of the 13.72 meV mode in both the low and high-incident energy datasets.

To ensure that all the crystal field parameters are real, we choose the local y-axis to be perpendicular to the mirror plane associated with the $C_s$ point group \cite{Hutchings_1964, Scheie_2018}. As mentioned above, this corresponds to the [110] crystallographic direction for one of the dimer sublattices and it is guaranteed to be one of the principal $g$-tensor directions by symmetry. On the other hand, the two principal $g$-tensor directions in the $xz$-plane are strongly dependent on the details of the surrounding ligands and they are not known a priori. Our choice of y-axis generates the following CEF Hamiltonian with 15 terms: 

\begin{equation*}\label{HamiltonianEr}
\begin{aligned}
\mathcal{H}_{CEF} = &B^0_2\hat{O}^0_2 + B^1_2\hat{O}^1_2 + B^2_2\hat{O}^2_2 + B^0_4\hat{O}^0_4 + B^1_4\hat{O}^1_4 + B^2_4\hat{O}^2_4\\
+ &B^3_4\hat{O}^3_4 + B^4_4\hat{O}^4_4 + B^0_6\hat{O}^0_6 + B^1_6\hat{O}^1_6 + B^2_6\hat{O}^2_6 + B^3_6\hat{O}^3_6 \\
+ &B^4_6\hat{O}^4_6 + B^5_6\hat{O}^5_6 +B^6_6\hat{O}^6_6. 
\end{aligned}
\end{equation*}
where $B_n^m$ and $\hat{O}_n^m$ represent the crystal field parameters and the corresponding operators in Stevens notation. Note that $B_n^m = \lambda_n^m \theta_n A_n^m$, where $\lambda^m_n$ are normalization constants \cite{21_scheie}, $\theta_n$ are constants associated with the electron orbitals of Er$^{3+}$ \cite{Stevens, Jensen_1991}, and $A_n^m$ are the crystal field parameters in Wybourne notation. 

The standard approach for determining crystal field parameters from INS data for higher point-symmetry systems is to perform a point charge calculation to obtain reasonable starting values and then refine the parameters through a least-squares fitting process with the data using the typical $\chi^2$ minimization function. For an under-constrained problem with a low point symmetry such as the present case, there is a strong likelihood that this procedure will not result in a meaningful solution since the fitting routine may get stuck in the same local minimum each time. To mitigate this issue and facilitate quick comparisons between INS data and crystal field models, the software package CrysFieldExplorer was developed recently \cite{MaCEF}. One key advantage of CrysFieldExplorer is the use of a revised cost/loss function for comparing the experimental and calculated CEF energy levels, which significantly reduces the likelihood of getting stuck in local minima. CrysFieldExplorer also takes advantage of high-performance parallel computing so rapid searches through complex parameter spaces can be conducted. This allows one to bypass the crystal field point charge calculation step in the determination of crystal field parameters from INS data and set up a random parameter initialization instead so that the parameter space search is not biased towards particular local minima. Note that CrysFieldExplorer reports crystal field parameters as $B_n^m/\theta_n$.

\setlength{\tabcolsep}{8pt}

\begin{table}[]

\caption{Measured CEF energy levels and the integrated intensities for all the transitions measured with INS. When the initial state corresponds to the CEF ground state doublet (i.e. 0), the data was collected at 5 K. Otherwise, the data was collected at 25~K. All the measured integrated intensities were normalized to the value for the 1.77~meV mode. The calculated integrated intensities for the two most probable CEF parameter sets discussed in the main text are also presented.}
\begin{tabular}{c l l l l l} 
\hline\hline
Transition &  $E_{obs}$(meV) & $I_{obs}$ & $I_{calc1}$ & $I_{calc2}$ \\
\hline
0 - 1      &     1.77(3)          & 1.0       & 1.0  & 1.0 \\
0 - 2      &     5.25(3)          & 0.365(35)       & 0.368 & 0.353 \\
0 - 3     &     7.17(5)          & 0.004(2)       & 0.017 & 0.036 \\
0 - 4     &     13.72(1)         & 0.167(5)       & 0.172 & 0.141\\
0 - 5     &     22.58(3)         & 0.074(4)       & 0.078 & 0.061 \\
0 - 6     &     27.81(6)         &  0.027(4)      & 0.029 & 0.031 \\
0 - 7     &     49.24(28)        & 0.010(2)       & 0.012 & 0.028 \\
1 - 2     &     3.44          & 0.207(22)      & 0.175      & 0.175 \\
2 - 4     &     8.29          & 0.039(3)       & 0.005      & 0.005 \\
1 - 4     &     11.81         & 0.023(3)       & 0.025      & 0.067      \\
    \hline\hline
    \end{tabular}
    \label{tableexp}
\end{table}

Promising crystal field parameter sets for Er$_2$Be$_2$SiO$_7$ were obtained by considering a custom loss function of the form:
\begin{equation}\label{LossFunction}
L_{tot} = L_{E} + L_{Int} + L_{M}
\end{equation}
where $L_E$, $L_{Int}$, and $L_M$ are the loss function components for the INS energy levels, the INS integrated intensities, and the 2~K powder-averaged $M$ vs $H$ data respectively. Their functional forms are given by:
\begin{equation}\label{LE}
L_{E} = log_{10}\left( \sum_i \dfrac{det\{ (E^{exp}_i +E^{cal}_0)\mathcal{I} - \mathcal{H}\}^2 }{det \{E^{exp}_i \mathcal{I}\}^2} \right),
\end{equation}
\begin{equation}\label{LInt}
L_{Int} = \dfrac{\sqrt{\sum (I^{exp}_i - I^{calc}_i)^2}}{\sqrt{\sum (I^{exp}_i)^2}},
\end{equation}
\begin{equation}\label{LM}
L_{M} = \dfrac{\sqrt{\sum (M^{exp}_i - M^{calc}_i)^2}}{\sqrt{\sum (M^{exp}_i)^2}},
\end{equation}
where $\mathcal{I}$ represents the identity matrix, $\mathcal{H}$ is the CEF Hamiltonian, and $det$ denotes the determinant of a matrix. While $L_E$ includes contributions from the 5~K INS data only, $L_{Int}$ includes intensities from both the 5~K and 25~K INS datasets. The summation on index $i$ in Eq.~\ref{LM} is carried out using a field interval three times larger than the measurement step size to reduce the computation time of the powder-averaged magnetization calculations.  

\begin{figure}
\centering
\scalebox{0.32}{\includegraphics{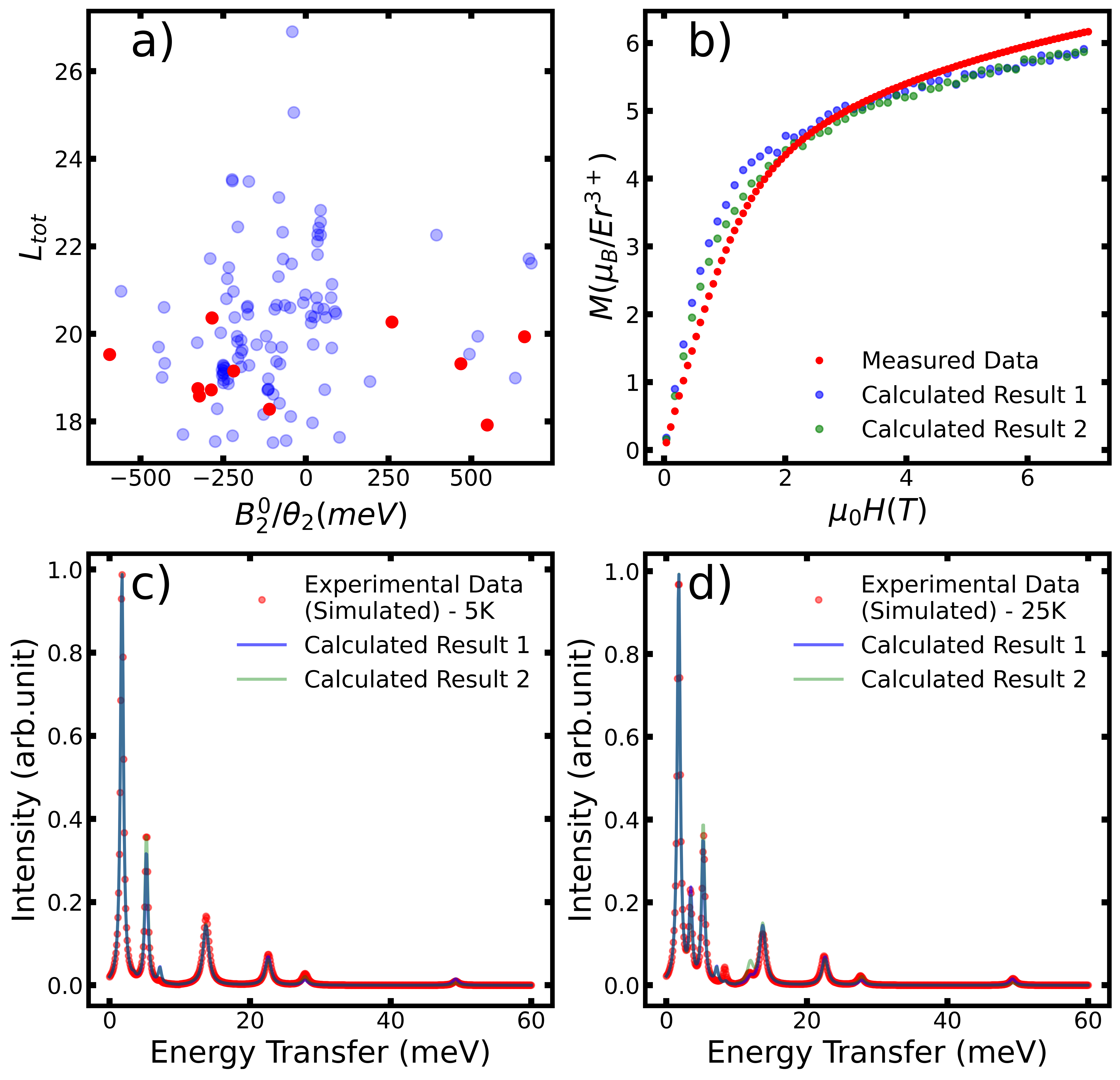}}
\caption{(a) The loss function value vs $B_2^0/\theta_2$ for all 116 crystal field parameter solutions determined by CrysFieldExplorer. The solid red circles represent the 12 solutions that satisfy the hard $g$-tensor constraints as discussed in the main text. (b) A comparison of the calculated (blue and green circles) vs measured (red circles) powder-averaged magnetization. The calculated magnetization curves for the two most probable crystal field solutions (as discussed in the main text) are shown. (c),(d) A comparison of the calculated (blue and green curves) vs simulated experimental INS spectra (red circles) at 5~K and 25~K respectively. The calculated spectra for the two most probable crystal field solutions are superimposed on the simulated data.  
}
\label{Fig6}
\end{figure}

 The complete set of INS energy levels and integrated intensities included in the fitting process are presented in Table~\ref{tableexp}. Due to the large crystal field parameter space, CrysFieldExplorer used the covariance matrix adaptation evolution strategy (CMA-ES) to minimize $L_{E} + L_{Int}$ initially. The fitting routine was run over 1600 times to sample the loss space and 116 solutions were found that matched the INS data significantly better than other solutions. Due to the computational cost of simulating powder-averaged magnetization data, $L_M$ was not included in the initial optimization but was later used in conjunction with the loss function from the INS data to assess the candidacy of the remaining 116 possible solutions. $L_{tot}$ vs $B_2^0/\theta_2$ for each of the 116 solutions is plotted in Fig.~\ref{Fig6}(a) with the chosen coordinate system for the crystal field parameters corresponding to the one with $B_2^1 =$~0, which has often been used for this low point symmetry previously to make comparisons between different crystal field parameters more straightforward \cite{08_gruber}. In other words, for the specific Er ion being considered here the local $y$-axis is aligned with the crystallographic [110] direction and the $x$ and $z$-axes make an arbitrary angle with the crystallographic [1-10] and [001] directions respectively. The large number of solutions with low $L_{tot}$ values indicate that a simultaneous fit of these three datasets is insufficient to fully constrain the 15 CEF parameters for this system.

To further constrain the possible crystal field parameter solutions, we use the 2~K $M$ vs $H$ data shown in Fig.~\ref{Fig4}(a) to establish approximate lower and upper bounds for the $g$-tensor components of the crystal field ground state doublet. We note that these bounds are based on the assumption that the principal $g$-tensor components are along the [110], [1-10], and [001] crystallographic directions for the Er ions forming one dimer sublattice, which correspond to the $y$, $x$, and $z$-axis in the local frame respectively. For the second Er dimer sublattice, the in-plane crystallographic directions corresponding to the $x$ and $y$ axes are reversed. Although the monoclinic site symmetry of the Er ions in this system ensures that the principal directions of the $g$-tensor in the $xz$-plane cannot be uniquely determined by symmetry constraints, a nearest-neighbor point charge calculation of the crystal field levels suggests that they are within 20$^\circ$ of our assumption. A magnetic structure refinement of recent low-temperature neutron diffraction data collected on the sister compound Dy$_2$Be$_2$SiO$_7$ provides further support for this assumption, as the moment direction was found to be in the $xz$-plane with the local $x$-axis (or true Ising axis) making an angle of $\sim$~17$^\circ$ with the crystallographic [1-10] (or [110]) direction \cite{brassington_unpublished}.

Subtracting off the linear van Vleck contribution for each field direction leads to saturation magnetization values of 2.65, 3.92, and 4.83~$\mu_B$ for $H \parallel$~[110], [100], and [001] respectively. As described above, the Er moments display a large anisotropy in the SSL planes, with a strong tendency to point along their local $y$-axis directions. This behavior suggests that $g_{xx}$ should be small. The other interesting observation from this data is that $M^{110}_{\rm sat}/M^{100}_{\rm sat} \approx$~1/$\sqrt{2}$ and $M^{110}_{\rm sat} /M^{001}_{\rm sat} \approx$~1/2. Since we argued above that only one Er dimer sublattice contributes to the magnetization with a magnetic field $H \parallel [110]$  while both sublattices should contribute equally when $H \parallel [001]$, this implies that $g_{yy}$ and $g_{zz}$ are very similar and the system has a quasi-XY magnetic anisotropy. As we will discuss in the next section, we also find that the ordered moment points along the local $y$ direction. Therefore, we assume that appropriate crystal field solutions should have $g_{yy} > g_{zz}$. Finally, we can place hard constraints on the maximum possible $g_{yy}$ and $g_{zz}$ values from the 2~K saturation magnetization (note that these are not exact $g_{yy}$ and $g_{zz}$ values due to the low-lying crystal field level at 1.77~meV) and the minimum possible $g_{yy}$ value from the magnitude of the ordered moment determined from neutron diffraction. We find that the upper bounds on $g_{yy}$ and $g_{zz}$ are given by 4$M^{110}_{sat} =$~10.6 and 2$M^{001}_{sat} =$~9.66 respectively. The magnitude of the ordered moment, as determined from neutron diffraction, has a value of 2.47(3)~$\mu_B$ and sets a lower limit on $g_{yy} =$~4.94. 

With the hard constraints on the $g$-tensor components of the crystal field ground state doublet now established as $4.94 < g_{yy} < 10.6$ and $0 < g_{zz} < 9.66$, the number of viable solutions from the CrysFieldExplorer analysis described above can be drastically reduced. Note that $g_{xx}$ and $g_{zz}$ are interchangeable here due to two degenerate sets of crystal field parameters related by a 90$^\circ$ rotation about the $y$-axis that we could not differentiate between when fitting the INS data above. Taking these criteria into account drastically reduces the number of acceptable solutions from 116 to 12. The remaining solutions that satisfy these $g$-factor criteria have been labeled with solid red circles in Fig.~\ref{Fig6}(a). Furthermore, only 10 of the remaining solutions are unique. The crystal field parameters, diagonal g-tensor components, and $L_{tot}$ values for each of these unique solutions are presented in Table~\ref{tableallsol} in order of increasing $L_{tot}$. The calculated integrated intensities for the $L_{tot} =$~17.92 ($I_{calc1}$) and 18.28 ($I_{calc2}$) solutions from Table~\ref{tableallsol} are shown in Table~\ref{tableexp}. Figure~\ref{Fig6}(b) compares the calculated $M$ vs $H$ to the experimental data for the same two solutions. Figures~\ref{Fig6}(c) and (d) show similar comparisons with the 5~K and 25~K INS data respectively. The experimental (simulated) neutron scattering spectra were plotted here by using a sum of Lorentzian functions with the measured (calculated) energy levels, the measured (calculated) integrated intensities, and fixed widths independent of energy transfer. Both solutions successfully account for most energy levels and their relative integrated intensities, although it is difficult to capture the small feature at $\sim$ 7 meV accurately as the low intensity ensures that its contribution to $L_{tot}$ is minimal. 

\setlength{\tabcolsep}{1.5pt}
\def\arraystretch{1.5}
\begin{table*}
    \caption{The crystal field parameters (in Stevens notation), the diagonal $g$-tensor components, and the $L_{tot}$ values for the ten unique solutions identified with CrysFieldExplorer that satisfy the hard $g$-tensor constraints discussed in the main text. The crystal field parameters are provided in units of meV.}
    \begin{tabular}{c| c c c c c c c c c c c c c c c c c c|c} 
    \hline\hline
        $B_n^m$& 1 &  2 & 3 & 4 & 5 & 6 & 7 & 8 & 9 & 10 \\
        \hline
$B_2^0$   & 1.40E-01 & -2.87E-02 & -9.40E-02 &-1.12E-01 & -5.56E-02 
 & 1.28E-01 & -1.53E-01 & 1.70E-01 & 1.58E-01 &-1.36E-01 \\
         
$B_2^1$ & 0 &  0 & 0 & 0 & 0 & 0 & 0 & 0 & 0 & 0 \\ 

$B_2^2$  & -6.20E-02 &  3.00E-01 & 1.85E-01 & 1.87E-01 & 2.01E-01 
 & 1.53E-02 & 1.76E-01 & -1.14E-01 & 2.30E-02 & 2.28E-01 \\
         
$B_4^0$ & 9.95E-04
 &  -1.51E-05
 & 4.20E-05
 & 1.29E-04
 & 2.34E-04
 & 2.43E-04
 & 5.28E-04
 & 6.04E-04
 & 2.71E-04
 & 1.59E-04 \\
      
$B_4^1$ & 3.87E-04
 &  1.53E-03
 & -1.71E-03
 & 4.80E-04
 & -2.78E-03
 & 6.04E-03
 & 6.30E-04
 & -2.18E-03
 & -5.15E-03
 & -1.22E-04 \\ 

$B_4^2$ & 2.57E-03
 &  -2.02E-03
 & -2.17E-03
 & -3.04E-03
 & -4.62E-04
 & -1.96E-04
 & -2.27E-03
 & -1.78E-03
 & -1.39E-03
 & -2.30E-03 \\

$B_4^3$ & -2.21E-03
 &  4.62E-03
 & 1.10E-02
 & -9.81E-04
 & 1.57E-02
 & 6.70E-03
 & 5.51E-03
 & -6.44E-03
 & 1.85E-03
 & 1.20E-02
 \\

$B_4^4 $ & -9.19E-04
 &  3.04E-03
 & 3.13E-03
 & 5.06E-03
 & 8.84E-04
 & 8.08E-04
 & 1.74E-03
 & 4.57E-04
 & 1.81E-03
 & 1.25E-03
 \\

$B_6^0$  & 8.98E-07
 &  2.11E-06
 & 3.64E-07
 & 7.51E-07
 & 1.23E-07
 & 4.68E-06
 & -7.31E-07
 & 1.73E-06
 & -7.24E-07
 & 7.85E-07
 \\

$B_6^1$  & 1.11E-05
 &  -3.10E-05
 & -6.44E-06
 & -8.30E-06
 & 1.48E-05
 & 5.63E-05
 & 6.95E-07
 & -1.13E-05
 & -3.68E-05
 & -1.31E-05
 \\
      
$B_6^2$  & 3.71E-06
 &  -2.86E-05
 & 2.21E-05
 & 1.59E-05
 & 1.44E-05
 & 1.25E-05
 & 1.22E-05
 & -1.86E-05
 & -4.28E-05
 & 2.86E-05
 \\
      
$B_6^3$  & 8.98E-06
 &  8.71E-05
 & -4.14E-05
 & 3.08E-05
 & -8.59E-05
 & 9.96E-06
 & -3.46E-05
 & 6.75E-05
 & 2.34E-05
 & -4.22E-05
 \\
      
$B_6^4$  & -3.50E-05
 & -5.73E-06
 & -2.69E-05
 & -1.81E-05
 & -1.72E-05
 & -3.10E-06
 & -8.13E-06
 & -2.13E-05
 & -1.23E-06
 & -2.71E-05
 \\
      
$B_6^5$  & 1.11E-04
 &  1.73E-06
 & 1.66E-04
 & 4.86E-05
 & 1.09E-04
 & -1.14E-04
 & -2.32E-04
 & 1.35E-04
 & 1.23E-04
 & -5.20E-05
 \\
      
$B_6^6$  & 1.94E-05
 &  4.66E-05
 & -5.49E-05
 & -4.91E-05
 & 4.49E-05
 & 1.18E-05
 & 3.71E-05
 & -1.79E-05
 & -1.51E-05
 & 4.43E-06
 \\
\hline

$g_{xx}$ & 2.18
 &  2.67
 & 5.83
 & 7.91
 & 7.46
 & 1.7
 & 2.26
 & 8.18
 & 6.98
 & 1.57
 \\

$g_{yy}$  & 10.1
 &  8.15
 & 10.3
 & 9.04
 & 9.73
 & 9.57
 & 9.01
 & 9.57
 & 10.5
 & 10.6
 \\

$g_{zz}$  & 6.16
 & 7.72
 & 2.43
 & 0.319
 & 1.66
 & 7.59
 & 8.55
 & 0.607
 & 1.59
 & 7.45
 \\
\hline

$L_{tot}$ & 17.9211
 &  18.2818
 & 18.5855
 & 18.7214
 & 19.156
 & 19.3198
 & 19.5293
 & 19.9351
 & 20.2746
 & 20.3626
 \\

    \hline\hline
    \end{tabular}
    \label{tableallsol}
\end{table*} 

\subsection{Zero-field Magnetic Ground State}

\begin{figure}
\centering
\scalebox{0.99}{\includegraphics{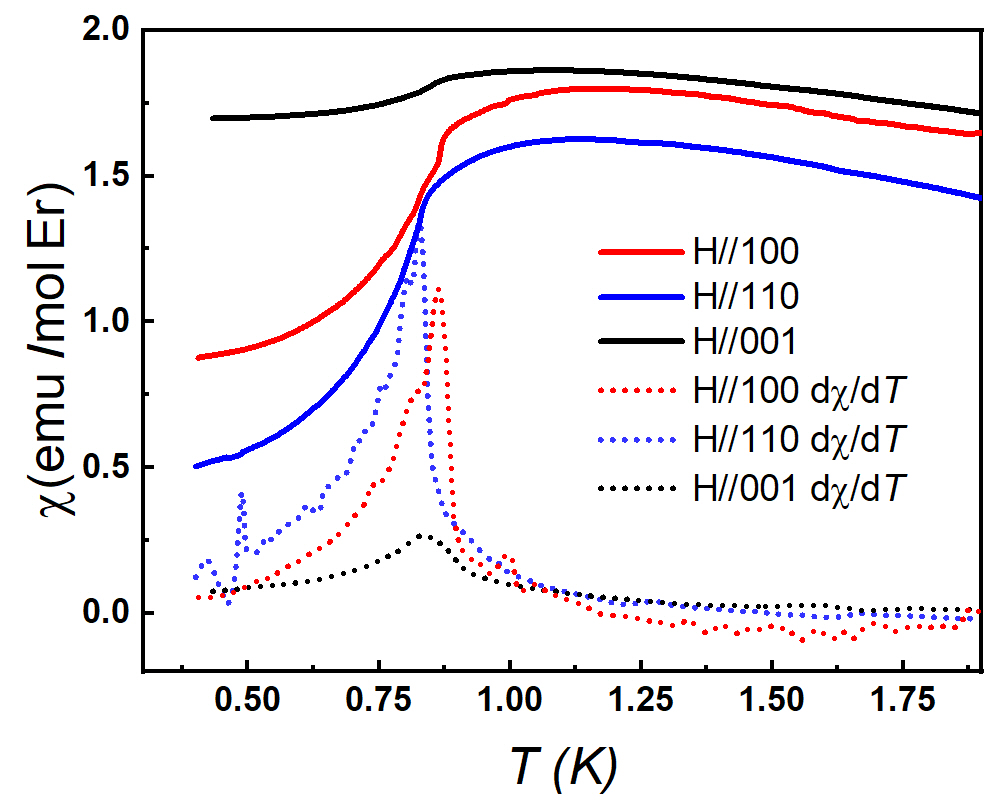}}
\caption{The DC susceptibility ($\chi$) and its derivative (d$\chi/dT$) vs $T$ measured at 0.1~T along three high-symmetry crystallographic directions. The peaks in (d$\chi/dT$), indicative of the magnetic transition temperature, are visible at 0.832, 0.827, and 0.865~K for applied fields along the [001], [110], and [100] directions respectively.}  
\label{Fig7}
\end{figure}

The magnetic susceptibility $\chi$ vs $T$ for a 0.1~T field applied along the crystallographic [001], [110], and [100] directions is shown in Fig.~\ref{Fig7}. All three datasets show a decrease in the signal at an average temperature of 0.841~K, which is indicative of a phase transition to long-range antiferromagnetic order. The enhanced signal reduction for the in-plane directions suggests that the ordered moments lie in the SSL plane. 

\begin{figure*}
\centering
\scalebox{0.6}{\includegraphics{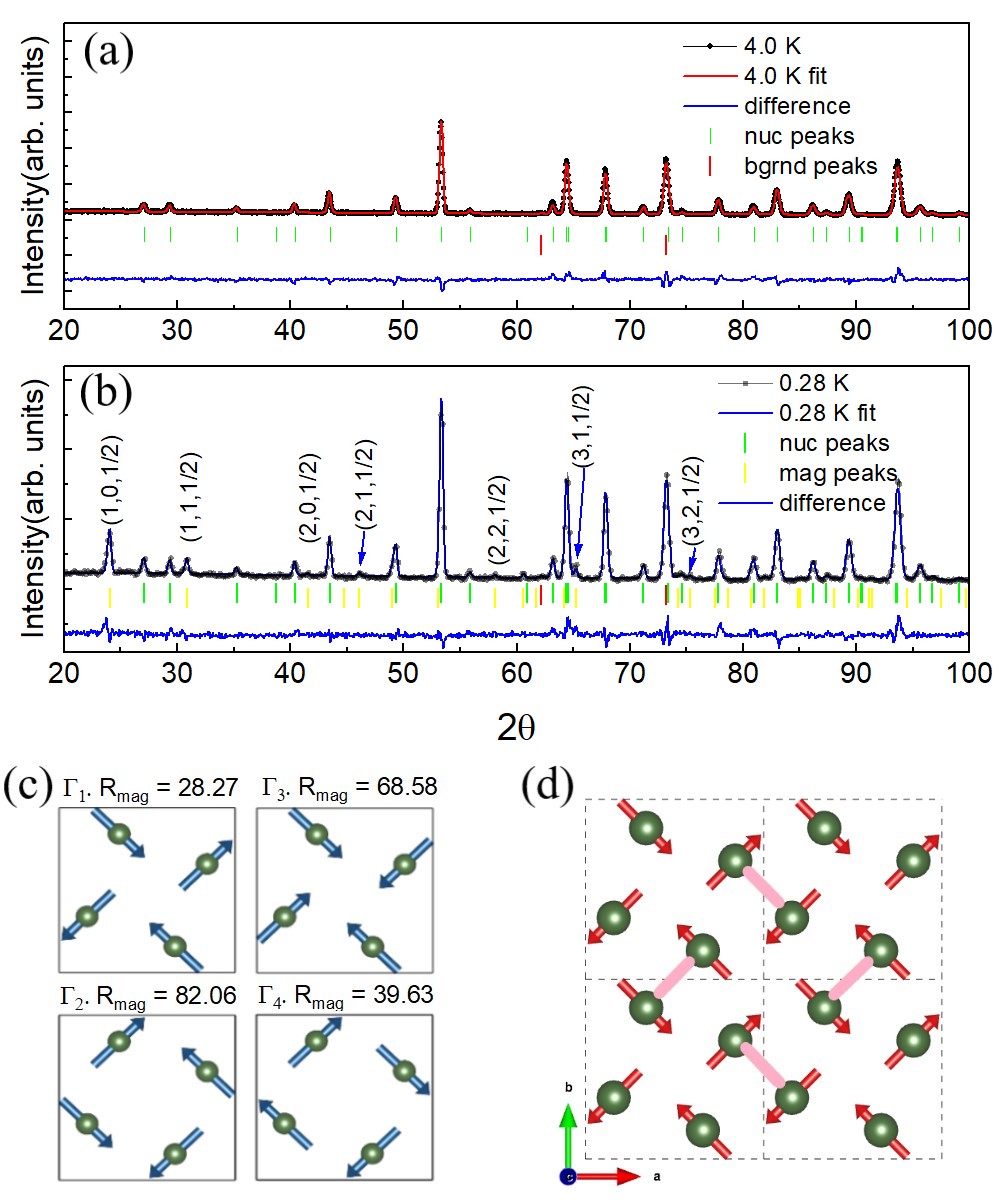}}
\caption{(a) The neutron powder diffraction pattern collected at 4~K (above $T_N$). The best Rietveld refinement result is superimposed on the data. The nuclear Bragg peaks from the sample and the background are indicated by green and red ticks respectively and the difference curve is shown below the Bragg peaks. (b) The neutron powder diffraction pattern collected at 0.28~K (below $T_N$). The best Rietveld refinement result, which now includes a magnetic phase, is again superimposed on the data. The magnetic Bragg peaks are denoted by the blue ticks. (c) Schematics of the in-plane spin-structures for the $\Gamma_1$ to $\Gamma_4$ models. The $R_{mag}$ values for the best refinements (which include out-of-plane components when allowed by symmetry) using each of these models are also shown. (d) The spin configuration of the $\Gamma_1$ magnetic structure realized by Er$_2$Be$_2$SiO$_7$ below $T_N$, with the dimer bonds shown in pink. The moments that form a square plaquette have a two-in-two-out structure and a magnitude of $2.47(3)~\mu_B$.  } 
\label{Fig8}
\end{figure*}

To determine the ordered spin configuration in the antiferromagnetic ground state, we performed neutron powder diffraction measurements. The nuclear structure was first confirmed by performing a Rietveld refinement using the FULLPROF software suite \cite{93_rodriguez} on a diffraction pattern collected above $T_N$ at 4~K. The data and the refinement results are shown in Fig.~\ref{Fig8}(a). The obtained lattice parameters, $a =$~7.262~\AA~and $c =$~4.744~\AA, agree well with refinement results from previous powder XRD data collected at 300~K \cite{Brassington_2024}. No evidence of a structural transition between 4~K and 300~K is found. 

A second neutron diffraction pattern was measured well below $T_N$ at 0.28~K. The nuclear phase was first refined using the parameters obtained from the 4~K fit as a starting point and no significant changes to the crystal structure were identified. Several additional Bragg peaks are visible in the low-temperature data and they could be indexed to a magnetic propagation vector of $\vec{k} =$~(0,0,0.5). Since there is only one SSL plane per chemical unit cell for the $R_2$Be$_2$SiO$_7$ family, this implies that the dominant interplane coupling is antiferromagnetic (AFM). With the magnetic propagation vector established, the software package SARAH \cite{WILLS_2006} was used to identify all five irreducible representations (IRs) in Kovalev's notation \cite{93_kovalev} allowed by symmetry. The $\Gamma_1$ - $\Gamma_4$ IRs correspond to spin states with AFM dimers, while the $\Gamma_5$ IR has ferromagnetic (FM) dimers. The $\Gamma_1$ and $\Gamma_3$ magnetic structures have their moments confined to the local directions perpendicular to the dimer bonds in the $ab$-plane, with $\Gamma_1$ and $\Gamma_3$ consisting of `two-in-two-out' and `all-in-all-out' square plaquette spin configurations. The $\Gamma_2$ and $\Gamma_4$ magnetic structures can have in-plane moments parallel to the dimer bonds, out-of-plane moments, or some combination of both. There are two main differences between them. First, the $\Gamma_4$ in-plane moments form a circulating square plaquette `current loop' with a definite handedness, but this is not true for $\Gamma_2$. Secondly, in a given SSL plane the $c$-axis moments are parallel for the $\Gamma_2$ structure, while they are anti-parallel for the two dimer sublattices of the $\Gamma_4$ configuration. Schematics for the in-plane moment configurations of the $\Gamma_1$ to $\Gamma_4$ magnetic structures are shown in Fig.~\ref{Fig8}(c).

\begin{figure*}
\centering
\scalebox{0.35}{\includegraphics{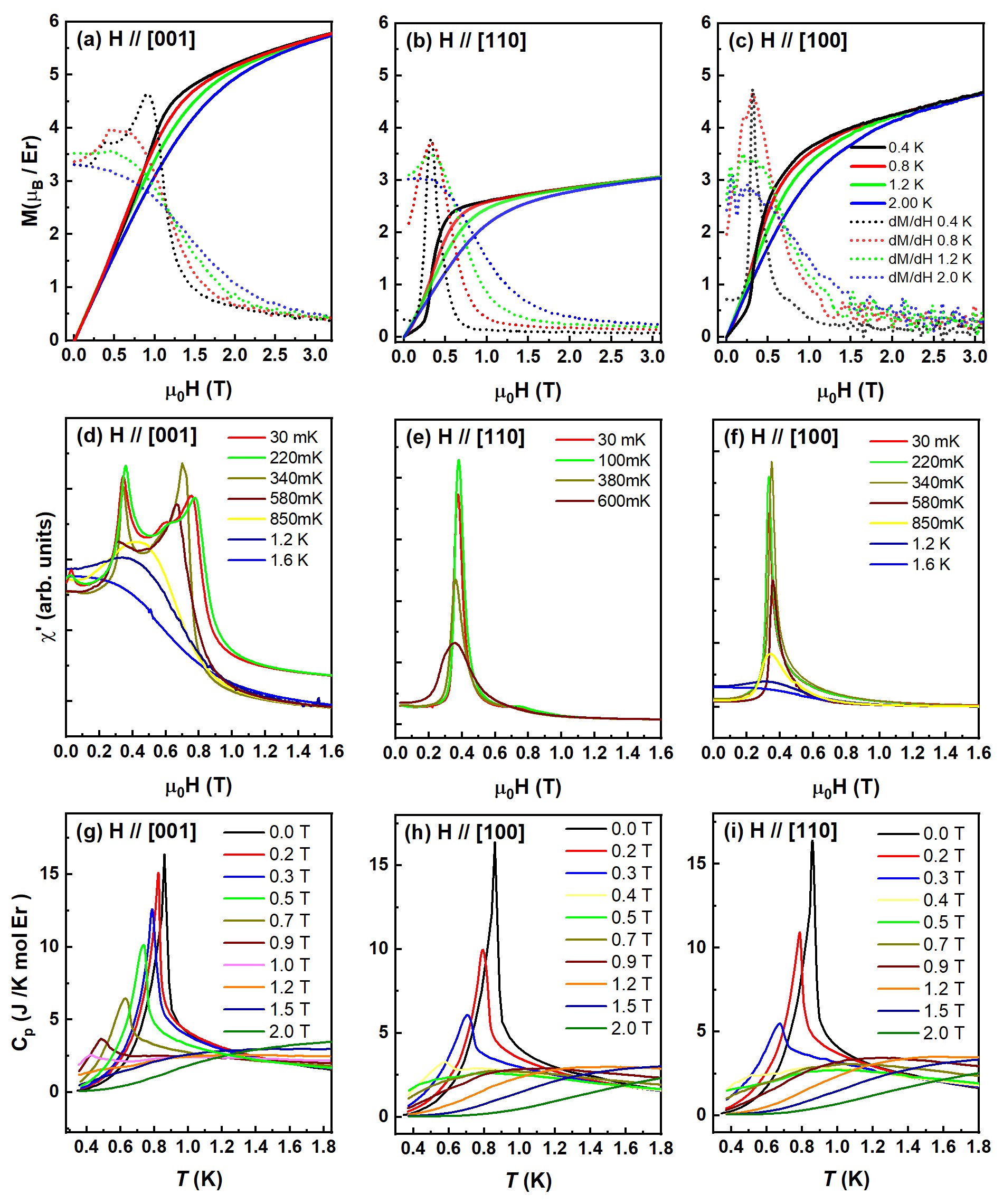}}
\caption{(a)-(c) DC magnetization and $dM/dH$ vs $H$ for applied magnetic fields along the [001], [110], and [100] directions at select temperatures. While only a single peak in $dM/dH$ indicative of one metamagnetic phase transition is visible in the low-$T$ in-plane field data, there are at least two $dM/dH$ peaks when $H \parallel$~[001]. (d)-(f) The real component of the AC susceptibility $\chi'$ vs $H$. While the results generally agree with the DC magnetization, there is an extra peak in the lowest-$T$ $H \parallel$~[001] $\chi'$ data which suggests that the $H$-$T$ phase diagram becomes more complex below 0.4~K. The appearance of peaks in the AC data at slightly lower fields, compared to the magnetization results, is likely due to a slight sample misalignment. (g)-(i) Heat capacity data $C_p$ vs $T$ for all three field orientations. In each case, the zero-field $\lambda$ anomaly indicative of magnetic order is suppressed with increasing field and replaced by a broad hump in the higher field range signifying the field-polarized phase regime. The onset of the polarized phase occurs at a much higher field for $H \parallel$~[001] due to the significant magnetic anisotropy of this system. }
\label{Fig9}
\end{figure*}

\begin{figure*}
\centering
\scalebox{1.05}{\includegraphics{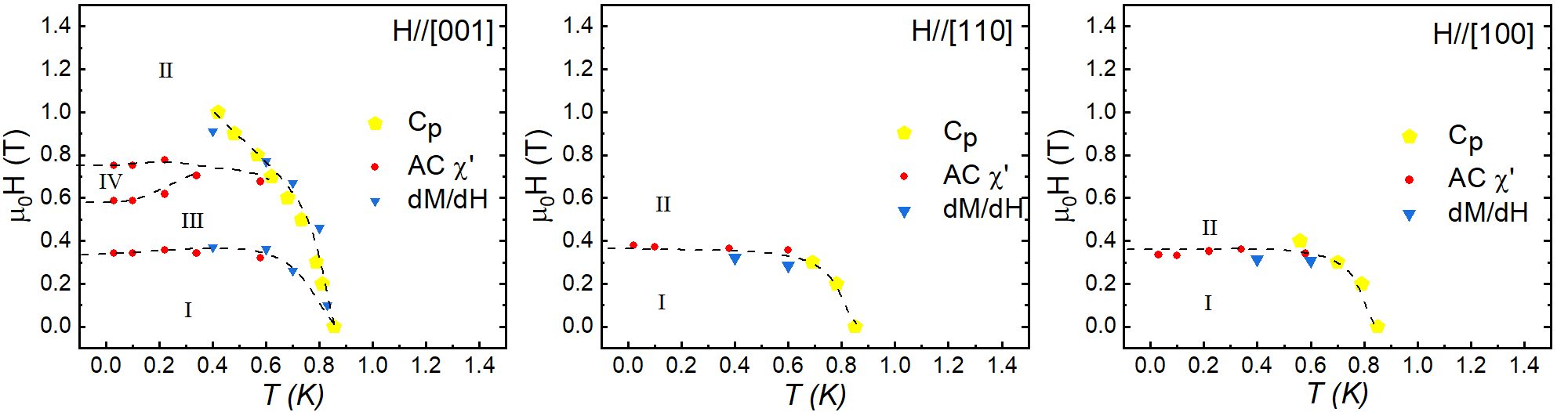}}
\caption{$H$-$T$ phase diagrams for Er$_2$Be$_2$SiO$_7$ with $H \parallel$~[001], $H \parallel$~[110], and $H \parallel$~[100]. Dashed lines are drawn as guides to the eye. The in-plane field behavior is quite simple and only consists of phase I and phase II, which correspond to the $\Gamma_1$ magnetic structure and the field-polarized state. There are two additional phases for $H \parallel$~[001] found at intermediate field ranges between phase I and II. Single crystal neutron diffraction suggests that phase III is a 2$\vec{k}$ magnetic structure corresponding to a field-induced canting of the moments towards the $c$-axis, while the nature of phase IV remains unknown.}
\label{Fig10}
\end{figure*}

For Er$_2$Be$_2$SiO$_7$, the negative Curie-Weiss temperatures identified above suggest that its low-temperature ground state consists of AFM dimers and therefore the most likely candidates are the $\Gamma_1$ - $\Gamma_4$ magnetic structures. To test this hypothesis, we performed systematic Rietveld refinements using the 0.28~K NPD data and the calculated magnetic diffraction patterns from each of the IRs in turn. Indeed, we find the largest $R_{mag} =$~141.7 for the $\Gamma_5$ model with the FM dimers. The fit quality improves substantially for all four AFM dimer models as shown in Fig.~\ref{Fig8}(c), but we find the best agreement ($R_{mag} =$~28.27) for the $\Gamma_1$ IR. Our $\Gamma_1$ magnetic structure refinement returns an ordered moment value of 2.47(3)~$\mu_B$ and indicates that all the Er moments point along the normal to their local mirror planes. A schematic of the refined magnetic structure viewed along the crystallographic $c$-axis is shown in Fig.~\ref{Fig8}(d).
 
\subsection{Magnetic Phase Diagrams}

We collected a variety of bulk characterization data with the applied magnetic field along the high-symmetry directions [001], [110], and [100] to explore the possibility of field-induced phase transitions and to establish $H$-$T$ magnetic phase diagrams. DC magnetization $M$ vs field $H$ is shown at selected temperatures for these three high-symmetry directions in Fig.~\ref{Fig9}(a)-(c). The first derivative $dM/dH$ is also shown on these same panels. Although magnetization plateaus are often observed in SSL systems, there is no evidence for them in this data. The in-plane field dependence of both $M$ and $dM/dH$ appears to be quite simple, with only one abrupt slope change in the 0.4~K magnetization data around 0.3-0.4~T corresponding to a large peak in $dM/dH$. This behavior is indicative of a single metamagnetic transition from the zero-field ground state established above to the field-polarized phase. The real part of the AC susceptibility, plotted in Fig.~\ref{Fig9}(e) and (f), shows that no additional features appear down to 30~mK. 

On the other hand, the $H \parallel$~[001] data appears to be more interesting. Aside from clear evidence in the 0.4~K magnetization data for a $\sim$~1.1~T phase transition to a field-polarized state, there is at least one lower-field peak in $dM/dH$ as shown in Fig.~\ref{Fig9}(a). Furthermore, the AC susceptibility data presented in Fig.~\ref{Fig9}(d) reveals that the $H$-$T$ phase diagram becomes more complex when the temperature is decreased below 0.4~K, as an additional peak appears in the data at an intermediate field. This feature may correspond to a third phase transition. The transition to the field-polarized phase occurs at a slightly lower field of 0.8 T in this data only, which is likely due to a slight sample misalignment.

Complementary heat capacity data with the field applied along the same three high-symmetry directions is shown in Figs.~\ref{Fig9}(g)-(i). All three datasets have the same general behavior, with clear $\lambda$ anomalies indicative of low-temperature magnetic order in the lower-field range and broader features characteristic of the field-polarized phase in the higher-field regime. As expected for the zero-field antiferromagnetic state, its ordering temperature is suppressed with increasing field applied along any direction. The field-evolution of the magnetic phase transition is extremely anisotropic for in-plane vs out-of-plane fields, with a much larger field required to reach saturation for $H \parallel$~[001]. 

Figure~\ref{Fig10} summarizes the bulk characterization results for the three high-symmetry directions. These $H$-$T$ phase diagrams were constructed from a combination of the heat capacity data ($\lambda$ anomaly temperature), the real part of the AC susceptibility (peak fields), and $dM/dH$ (peak fields). As described above, the phase diagrams for the two in-plane field directions are nearly indistinguishable and consist of a single metamagnetic transition from the low-field $\Gamma_1$ antiferromagnetic ground state to a field-polarized phase (see Figs.~\ref{Fig4}(b) and (c) for schematics of the high-field spin structures). The phase diagram for $H \parallel$~[001] is more complex, as we have identified two intermediate field-induced states.

\begin{figure}
\centering
\scalebox{0.36}{\includegraphics{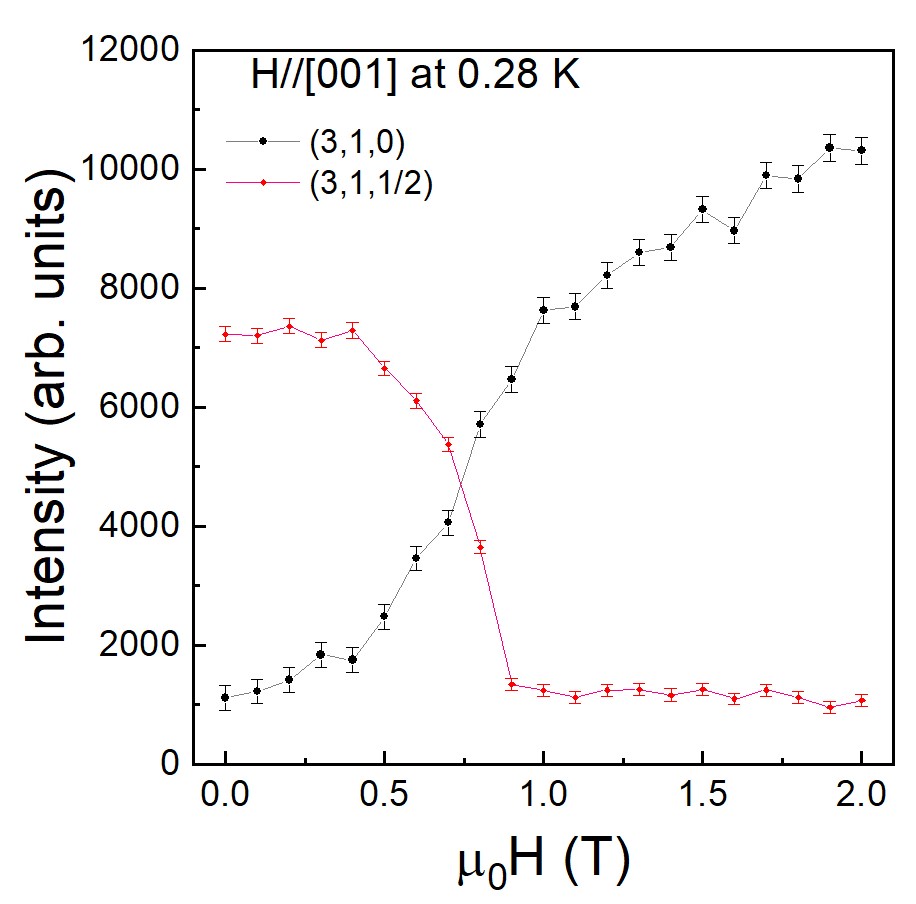}}
\caption{Field-dependence of the Bragg peak intensity for the (3,1,0) and (3,1,0.5) positions. While the (3,1,0.5) intensity drops sharply between 0.4 and 1~T, the (3,1,0) intensity tracks the field-dependence of the DC magnetization at comparable temperatures quite well. The field-dependence of these two peaks matches expectations for a field-induced canting of the moments towards the $c$-axis that cannot proceed until a small energy barrier due to the magnetic anisotropy of the system is overcome first.}
\label{Fig11}
\end{figure}

To investigate the microscopic origin of these phases in more detail, we performed a single crystal neutron diffraction experiment with the magnetic field applied along the [001] direction. This was a challenging measurement due to the low-field magnetic propagation vector of $\vec{k} =$~(0,0,0.5), which can only be probed with this field direction using a horizontal-field cryomagnet or a vertical-field cryomagnet with out-of-plane detector geometry. We performed this experiment on DEMAND since it offers the latter combination, although very few magnetic Bragg peaks could be measured in each phase due to the limited vertical acceptance of the cryomagnet. 

The main results of the neutron diffraction experiment are shown in Fig.~\ref{Fig11}. Here, we present the field-evolution of the (3,1,0) and (3,1,0.5) Bragg peaks at a fixed temperature of 0.28~K. There are three different field regimes. For the lowest fields, the (3,1,0) peak intensity shows a modest increase while the (3,1,0.5) peak intensity is constant. For intermediate fields, the (3,1,0.5) peak and the (3,1,0) peak show a large decrease and increase respectively. For the highest fields, the (3,1,0.5) peak intensity is completely suppressed while the (3,1,0) peak shows another modest intensity increase. The field-evolution of the (3,1,0) peak tracks the 0.4~K magnetization data well, which shows that it is an effective probe of the net moment along the $c$-axis. These results indicate that one of the intermediate phases corresponds to a 2$\vec{k}$-structure, with the in-plane and out-of-plane moment components giving rise to the $\vec{k} =$~(0,0,0.5) and $\vec{k} =$~0 magnetic propagation vectors respectively. The onset of the polarized phase corresponds to the disappearance of the (3,1,0.5) peak and hence the complete suppression of the in-plane moment component. There appears to be a weak energy barrier that must be overcome by the magnetic field to reach the $2\vec{k}$-state since the (3,1,0.5) peak intensity does not decrease immediately after the field is applied. This behavior provides further evidence that the Er moments in this system have quasi-XY rather than true XY magnetic anisotropy. Finally, we note that there is no evidence for the second intermediate-field phase in this data, which may be due to the elevated base temperature as compared to the AC susceptibility measurement.

\section{Conclusions}

We investigated the structural and magnetic properties of both polycrystalline and single crystal samples of the Shastry-Sutherland lattice system Er$_2$Be$_2$SiO$_7$ using a variety of bulk characterization and scattering techniques. This material crystallizes in the tetragonal P-4$2_1$m space group (113) over a wide temperature range from 0.28~K to 300~K. The Er moments have a quasi-XY magnetic anisotropy with a small preference to lie along the normal to their local mirror plane, which corresponds to the crystallographic [110] or [1-10] direction for the two orthogonal dimer sublattices. Inelastic neutron scattering data shows that the $J_{\rm eff} =$~1/2 model is not appropriate for this system due to a low-lying crystal field level and it was also used to establish some probable crystal field parameters for this material. 

Er$_2$Be$_2$SiO$_7$ exhibits long-range non-collinear magnetic order below $T_N =$~0.841~K. The $\Gamma_1$ structure consists of antiferromagnetic dimers with in-plane moments perpendicular to the dimer bond, a two-in-two-out spin structure on the square plaquettes that connect four different dimers together, and antiferromagnetic coupling between Shastry-Sutherland planes. A modest in-plane magnetic field $\sim$~0.4~T induces a magnetic transition from the $\Gamma_1$ magnetic structure to a field-polarized phase. An $H \parallel$~[001] field generates a more complicated $H$-$T$ phase diagram with at least two intermediate field states between the $\Gamma_1$ and field-polarized phases. One intermediate phase has a 2$\vec{k}$ magnetic structure corresponding to a gradual canting of the moments towards the $c$-axis with increasing field, while the microscopic origin of the second intermediate phase remains unknown. 

We expect our work to motivate comprehensive magnetism studies of other isostructural $R_2$Be$_2$SiO$_7$ and $R_2$Be$_2$GeO$_7$ family members. There is a strong likelihood that the rare-earth-based Shastry-Sutherland planes, combined with specific magnetic anisotropies and insulating behavior, will generate exotic zero-field magnetic ground states and field-induced magnetization plateaus that can be readily compared to theoretical predictions of the beyond-Heisenberg Shastry-Sutherland model.

\section{Acknowledgments}

Research at the University of Tennessee is supported by the National Science Foundation, Division of Materials Research under Award No. NSF-DMR-2003117. The work at Michigan State University is supported by the U.S.DOE-BES under Contract DE-SC0023648. The work performed at NHMFL is supported by the NSF Cooperative Agreement No.DMR-1644779 and the State of Florida. A portion of this research used resources at the Spallation Neutron Source and the High Flux Isotope Reactor, which are DOE Office of Science User Facilities operated by Oak Ridge National Laboratory.

\bibliography{Er2Be2SiO7_11202023}

\begin{thebibliography}{86}%
\makeatletter
\providecommand \@ifxundefined [1]{%
 \@ifx{#1\undefined}
}%
\providecommand \@ifnum [1]{%
 \ifnum #1\expandafter \@firstoftwo
 \else \expandafter \@secondoftwo
 \fi
}%
\providecommand \@ifx [1]{%
 \ifx #1\expandafter \@firstoftwo
 \else \expandafter \@secondoftwo
 \fi
}%
\providecommand \natexlab [1]{#1}%
\providecommand \enquote  [1]{``#1''}%
\providecommand \bibnamefont  [1]{#1}%
\providecommand \bibfnamefont [1]{#1}%
\providecommand \citenamefont [1]{#1}%
\providecommand \href@noop [0]{\@secondoftwo}%
\providecommand \href [0]{\begingroup \@sanitize@url \@href}%
\providecommand \@href[1]{\@@startlink{#1}\@@href}%
\providecommand \@@href[1]{\endgroup#1\@@endlink}%
\providecommand \@sanitize@url [0]{\catcode `\\12\catcode `\$12\catcode `\&12\catcode `\#12\catcode `\^12\catcode `\_12\catcode `\%12\relax}%
\providecommand \@@startlink[1]{}%
\providecommand \@@endlink[0]{}%
\providecommand \url  [0]{\begingroup\@sanitize@url \@url }%
\providecommand \@url [1]{\endgroup\@href {#1}{\urlprefix }}%
\providecommand \urlprefix  [0]{URL }%
\providecommand \Eprint [0]{\href }%
\providecommand \doibase [0]{https://doi.org/}%
\providecommand \selectlanguage [0]{\@gobble}%
\providecommand \bibinfo  [0]{\@secondoftwo}%
\providecommand \bibfield  [0]{\@secondoftwo}%
\providecommand \translation [1]{[#1]}%
\providecommand \BibitemOpen [0]{}%
\providecommand \bibitemStop [0]{}%
\providecommand \bibitemNoStop [0]{.\EOS\space}%
\providecommand \EOS [0]{\spacefactor3000\relax}%
\providecommand \BibitemShut  [1]{\csname bibitem#1\endcsname}%
\let\auto@bib@innerbib\@empty
\bibitem [{\citenamefont {Ramirez}(1994)}]{94_ramirez}%
  \BibitemOpen
  \bibfield  {author} {\bibinfo {author} {\bibfnamefont {A.~P.}\ \bibnamefont {Ramirez}},\ }\bibfield  {title} {\bibinfo {title} {Strongly geometrically frustrated magnets},\ }\href@noop {} {\bibfield  {journal} {\bibinfo  {journal} {Annual Review of Materials Research}\ }\textbf {\bibinfo {volume} {24}},\ \bibinfo {pages} {453} (\bibinfo {year} {1994})}\BibitemShut {NoStop}%
\bibitem [{\citenamefont {Greedan}(2001)}]{00_greedan}%
  \BibitemOpen
  \bibfield  {author} {\bibinfo {author} {\bibfnamefont {J.~E.}\ \bibnamefont {Greedan}},\ }\bibfield  {title} {\bibinfo {title} {Geometrically frustrated magnetic materials},\ }\href@noop {} {\bibfield  {journal} {\bibinfo  {journal} {J. Mater. Chem.}\ }\textbf {\bibinfo {volume} {11}},\ \bibinfo {pages} {37} (\bibinfo {year} {2001})}\BibitemShut {NoStop}%
\bibitem [{\citenamefont {Moessner}\ and\ \citenamefont {Ramirez}(2006)}]{06_moessner}%
  \BibitemOpen
  \bibfield  {author} {\bibinfo {author} {\bibfnamefont {R.}~\bibnamefont {Moessner}}\ and\ \bibinfo {author} {\bibfnamefont {A.~P.}\ \bibnamefont {Ramirez}},\ }\bibfield  {title} {\bibinfo {title} {{Geometrical frustration}},\ }\href@noop {} {\bibfield  {journal} {\bibinfo  {journal} {Physics Today}\ }\textbf {\bibinfo {volume} {59}},\ \bibinfo {pages} {24} (\bibinfo {year} {2006})}\BibitemShut {NoStop}%
\bibitem [{\citenamefont {Mirebeau}\ and\ \citenamefont {Petit}(2014)}]{14_mirebeau}%
  \BibitemOpen
  \bibfield  {author} {\bibinfo {author} {\bibfnamefont {I.}~\bibnamefont {Mirebeau}}\ and\ \bibinfo {author} {\bibfnamefont {S.}~\bibnamefont {Petit}},\ }\bibfield  {title} {\bibinfo {title} {Magnetic frustration probed by inelastic neutron scattering: Recent examples},\ }\href@noop {} {\bibfield  {journal} {\bibinfo  {journal} {Journal of Magnetism and Magnetic Materials}\ }\textbf {\bibinfo {volume} {350}},\ \bibinfo {pages} {209} (\bibinfo {year} {2014})}\BibitemShut {NoStop}%
\bibitem [{\citenamefont {Balents}(2010)}]{10_balents}%
  \BibitemOpen
  \bibfield  {author} {\bibinfo {author} {\bibfnamefont {L.}~\bibnamefont {Balents}},\ }\href@noop {} {\bibfield  {journal} {\bibinfo  {journal} {Nature}\ }\textbf {\bibinfo {volume} {464}},\ \bibinfo {pages} {199} (\bibinfo {year} {2010})}\BibitemShut {NoStop}%
\bibitem [{\citenamefont {Norman}(2016)}]{16_norman}%
  \BibitemOpen
  \bibfield  {author} {\bibinfo {author} {\bibfnamefont {M.~R.}\ \bibnamefont {Norman}},\ }\href@noop {} {\bibfield  {journal} {\bibinfo  {journal} {Rev. Mod. Phys.}\ }\textbf {\bibinfo {volume} {88}},\ \bibinfo {pages} {041002} (\bibinfo {year} {2016})}\BibitemShut {NoStop}%
\bibitem [{\citenamefont {Savary}\ and\ \citenamefont {Balents}(2017)}]{17_savary}%
  \BibitemOpen
  \bibfield  {author} {\bibinfo {author} {\bibfnamefont {L.}~\bibnamefont {Savary}}\ and\ \bibinfo {author} {\bibfnamefont {L.}~\bibnamefont {Balents}},\ }\href@noop {} {\bibfield  {journal} {\bibinfo  {journal} {Rep. Prog. Phys.}\ }\textbf {\bibinfo {volume} {80}},\ \bibinfo {pages} {016502} (\bibinfo {year} {2017})}\BibitemShut {NoStop}%
\bibitem [{\citenamefont {Zhou}\ \emph {et~al.}(2017)\citenamefont {Zhou}, \citenamefont {Kanoda},\ and\ \citenamefont {Ng}}]{17_zhou}%
  \BibitemOpen
  \bibfield  {author} {\bibinfo {author} {\bibfnamefont {Y.}~\bibnamefont {Zhou}}, \bibinfo {author} {\bibfnamefont {K.}~\bibnamefont {Kanoda}},\ and\ \bibinfo {author} {\bibfnamefont {T.~K.}\ \bibnamefont {Ng}},\ }\bibfield  {title} {\bibinfo {title} {Quantum spin liquid states},\ }\href@noop {} {\bibfield  {journal} {\bibinfo  {journal} {Rev. Mod. Phys.}\ }\textbf {\bibinfo {volume} {89}},\ \bibinfo {pages} {025003} (\bibinfo {year} {2017})}\BibitemShut {NoStop}%
\bibitem [{\citenamefont {Knolle}\ and\ \citenamefont {Moessner}(2019)}]{19_knolle}%
  \BibitemOpen
  \bibfield  {author} {\bibinfo {author} {\bibfnamefont {J.}~\bibnamefont {Knolle}}\ and\ \bibinfo {author} {\bibfnamefont {R.}~\bibnamefont {Moessner}},\ }\bibfield  {title} {\bibinfo {title} {A field guide to spin liquids},\ }\href@noop {} {\bibfield  {journal} {\bibinfo  {journal} {Annu. Rev. Condens. Matter Phys.}\ }\textbf {\bibinfo {volume} {10}},\ \bibinfo {pages} {451} (\bibinfo {year} {2019})}\BibitemShut {NoStop}%
\bibitem [{\citenamefont {Takagi}\ \emph {et~al.}(2019)\citenamefont {Takagi}, \citenamefont {Takayama}, \citenamefont {Jackeli}, \citenamefont {Khaliullin},\ and\ \citenamefont {Nagler}}]{19_takagi}%
  \BibitemOpen
  \bibfield  {author} {\bibinfo {author} {\bibfnamefont {H.}~\bibnamefont {Takagi}}, \bibinfo {author} {\bibfnamefont {T.}~\bibnamefont {Takayama}}, \bibinfo {author} {\bibfnamefont {G.}~\bibnamefont {Jackeli}}, \bibinfo {author} {\bibfnamefont {G.}~\bibnamefont {Khaliullin}},\ and\ \bibinfo {author} {\bibfnamefont {S.~E.}\ \bibnamefont {Nagler}},\ }\href@noop {} {\bibfield  {journal} {\bibinfo  {journal} {Nat. Rev. Phys.}\ }\textbf {\bibinfo {volume} {1}},\ \bibinfo {pages} {264} (\bibinfo {year} {2019})}\BibitemShut {NoStop}%
\bibitem [{\citenamefont {Wen}\ \emph {et~al.}(2019)\citenamefont {Wen}, \citenamefont {Yu}, \citenamefont {Li}, \citenamefont {Yu},\ and\ \citenamefont {Li}}]{19_wen}%
  \BibitemOpen
  \bibfield  {author} {\bibinfo {author} {\bibfnamefont {J.}~\bibnamefont {Wen}}, \bibinfo {author} {\bibfnamefont {S.~L.}\ \bibnamefont {Yu}}, \bibinfo {author} {\bibfnamefont {S.}~\bibnamefont {Li}}, \bibinfo {author} {\bibfnamefont {W.}~\bibnamefont {Yu}},\ and\ \bibinfo {author} {\bibfnamefont {J.~X.}\ \bibnamefont {Li}},\ }\href@noop {} {\bibfield  {journal} {\bibinfo  {journal} {npj Quantum Mater.}\ }\textbf {\bibinfo {volume} {4}},\ \bibinfo {pages} {12} (\bibinfo {year} {2019})}\BibitemShut {NoStop}%
\bibitem [{\citenamefont {Broholm}\ \emph {et~al.}(2020)\citenamefont {Broholm}, \citenamefont {Cava}, \citenamefont {Kivelson}, \citenamefont {Nocera}, \citenamefont {Norman},\ and\ \citenamefont {Senthil}}]{20_broholm}%
  \BibitemOpen
  \bibfield  {author} {\bibinfo {author} {\bibfnamefont {C.}~\bibnamefont {Broholm}}, \bibinfo {author} {\bibfnamefont {R.~J.}\ \bibnamefont {Cava}}, \bibinfo {author} {\bibfnamefont {S.~A.}\ \bibnamefont {Kivelson}}, \bibinfo {author} {\bibfnamefont {D.~G.}\ \bibnamefont {Nocera}}, \bibinfo {author} {\bibfnamefont {M.~R.}\ \bibnamefont {Norman}},\ and\ \bibinfo {author} {\bibfnamefont {T.}~\bibnamefont {Senthil}},\ }\href@noop {} {\bibfield  {journal} {\bibinfo  {journal} {Science}\ }\textbf {\bibinfo {volume} {367}},\ \bibinfo {pages} {263} (\bibinfo {year} {2020})}\BibitemShut {NoStop}%
\bibitem [{\citenamefont {Chamorro}\ \emph {et~al.}(2021)\citenamefont {Chamorro}, \citenamefont {McQueen},\ and\ \citenamefont {Tran}}]{21_chamorro}%
  \BibitemOpen
  \bibfield  {author} {\bibinfo {author} {\bibfnamefont {J.~R.}\ \bibnamefont {Chamorro}}, \bibinfo {author} {\bibfnamefont {T.~M.}\ \bibnamefont {McQueen}},\ and\ \bibinfo {author} {\bibfnamefont {T.~T.}\ \bibnamefont {Tran}},\ }\href@noop {} {\bibfield  {journal} {\bibinfo  {journal} {Chem. Rev.}\ }\textbf {\bibinfo {volume} {121}},\ \bibinfo {pages} {2898} (\bibinfo {year} {2021})}\BibitemShut {NoStop}%
\bibitem [{\citenamefont {Yin}\ \emph {et~al.}(2022)\citenamefont {Yin}, \citenamefont {Lian},\ and\ \citenamefont {Hasan}}]{22_yin}%
  \BibitemOpen
  \bibfield  {author} {\bibinfo {author} {\bibfnamefont {J.-X.}\ \bibnamefont {Yin}}, \bibinfo {author} {\bibfnamefont {B.}~\bibnamefont {Lian}},\ and\ \bibinfo {author} {\bibfnamefont {M.~Z.}\ \bibnamefont {Hasan}},\ }\bibfield  {title} {\bibinfo {title} {Topological kagome magnets and superconductors},\ }\href@noop {} {\bibfield  {journal} {\bibinfo  {journal} {Nature}\ }\textbf {\bibinfo {volume} {612}},\ \bibinfo {pages} {647} (\bibinfo {year} {2022})}\BibitemShut {NoStop}%
\bibitem [{\citenamefont {Greedan}(2006)}]{06_greedan}%
  \BibitemOpen
  \bibfield  {author} {\bibinfo {author} {\bibfnamefont {J.~E.}\ \bibnamefont {Greedan}},\ }\bibfield  {title} {\bibinfo {title} {Frustrated rare earth magnetism: Spin glasses, spin liquids and spin ices in pyrochlore oxides},\ }\href@noop {} {\bibfield  {journal} {\bibinfo  {journal} {Journal of Alloys and Compounds}\ }\textbf {\bibinfo {volume} {408-412}},\ \bibinfo {pages} {444} (\bibinfo {year} {2006})},\ \bibinfo {note} {proceedings of Rare Earths'04 in Nara, Japan}\BibitemShut {NoStop}%
\bibitem [{\citenamefont {Gardner}\ \emph {et~al.}(2010)\citenamefont {Gardner}, \citenamefont {Gingras},\ and\ \citenamefont {Greedan}}]{10_gardner}%
  \BibitemOpen
  \bibfield  {author} {\bibinfo {author} {\bibfnamefont {J.~S.}\ \bibnamefont {Gardner}}, \bibinfo {author} {\bibfnamefont {M.~J.~P.}\ \bibnamefont {Gingras}},\ and\ \bibinfo {author} {\bibfnamefont {J.~E.}\ \bibnamefont {Greedan}},\ }\bibfield  {title} {\bibinfo {title} {Magnetic pyrochlore oxides},\ }\href@noop {} {\bibfield  {journal} {\bibinfo  {journal} {Rev. Mod. Phys.}\ }\textbf {\bibinfo {volume} {82}},\ \bibinfo {pages} {53} (\bibinfo {year} {2010})}\BibitemShut {NoStop}%
\bibitem [{\citenamefont {Starykh}(2015)}]{15_starykh}%
  \BibitemOpen
  \bibfield  {author} {\bibinfo {author} {\bibfnamefont {O.~A.}\ \bibnamefont {Starykh}},\ }\bibfield  {title} {\bibinfo {title} {Unusual ordered phases of highly frustrated magnets: a review},\ }\href@noop {} {\bibfield  {journal} {\bibinfo  {journal} {Reports on Progress in Physics}\ }\textbf {\bibinfo {volume} {78}},\ \bibinfo {pages} {052502} (\bibinfo {year} {2015})}\BibitemShut {NoStop}%
\bibitem [{\citenamefont {{Sriram Shastry}}\ and\ \citenamefont {Sutherland}(1981)}]{Shastry_1981}%
  \BibitemOpen
  \bibfield  {author} {\bibinfo {author} {\bibfnamefont {B.}~\bibnamefont {{Sriram Shastry}}}\ and\ \bibinfo {author} {\bibfnamefont {B.}~\bibnamefont {Sutherland}},\ }\bibfield  {title} {\bibinfo {title} {Exact ground state of a quantum mechanical antiferromagnet},\ }\href {https://www.sciencedirect.com/science/article/pii/037843638190838X} {\bibfield  {journal} {\bibinfo  {journal} {Physica B+C}\ }\textbf {\bibinfo {volume} {108}},\ \bibinfo {pages} {1069} (\bibinfo {year} {1981})}\BibitemShut {NoStop}%
\bibitem [{\citenamefont {Miyahara}\ and\ \citenamefont {Ueda}(1999)}]{Miyahara_1999}%
  \BibitemOpen
  \bibfield  {author} {\bibinfo {author} {\bibfnamefont {S.}~\bibnamefont {Miyahara}}\ and\ \bibinfo {author} {\bibfnamefont {K.}~\bibnamefont {Ueda}},\ }\bibfield  {title} {\bibinfo {title} {Exact dimer ground state of the two dimensional heisenberg spin system {SrCu$_2$(BO$_3$)$_2$}},\ }\href {https://doi.org/10.1103/PhysRevLett.82.3701} {\bibfield  {journal} {\bibinfo  {journal} {Phys. Rev. Lett.}\ }\textbf {\bibinfo {volume} {82}},\ \bibinfo {pages} {3701} (\bibinfo {year} {1999})}\BibitemShut {NoStop}%
\bibitem [{\citenamefont {Kageyama}\ \emph {et~al.}(1999)\citenamefont {Kageyama}, \citenamefont {Yoshimura}, \citenamefont {Stern}, \citenamefont {Mushnikov}, \citenamefont {Onizuka}, \citenamefont {Kato}, \citenamefont {Kosuge}, \citenamefont {Slichter}, \citenamefont {Goto},\ and\ \citenamefont {Ueda}}]{Kageyama_1999}%
  \BibitemOpen
  \bibfield  {author} {\bibinfo {author} {\bibfnamefont {H.}~\bibnamefont {Kageyama}}, \bibinfo {author} {\bibfnamefont {K.}~\bibnamefont {Yoshimura}}, \bibinfo {author} {\bibfnamefont {R.}~\bibnamefont {Stern}}, \bibinfo {author} {\bibfnamefont {N.~V.}\ \bibnamefont {Mushnikov}}, \bibinfo {author} {\bibfnamefont {K.}~\bibnamefont {Onizuka}}, \bibinfo {author} {\bibfnamefont {M.}~\bibnamefont {Kato}}, \bibinfo {author} {\bibfnamefont {K.}~\bibnamefont {Kosuge}}, \bibinfo {author} {\bibfnamefont {C.~P.}\ \bibnamefont {Slichter}}, \bibinfo {author} {\bibfnamefont {T.}~\bibnamefont {Goto}},\ and\ \bibinfo {author} {\bibfnamefont {Y.}~\bibnamefont {Ueda}},\ }\bibfield  {title} {\bibinfo {title} {Exact dimer ground state and quantized magnetization plateaus in the two-dimensional spin system {SrCu$_2$(BO$_3$)$_2$}},\ }\href@noop {} {\bibfield  {journal} {\bibinfo  {journal} {Phys. Rev. Lett.}\ }\textbf {\bibinfo {volume} {82}},\ \bibinfo {pages} {3168} (\bibinfo {year} {1999})}\BibitemShut {NoStop}%
\bibitem [{\citenamefont {Guo}\ \emph {et~al.}(2020)\citenamefont {Guo}, \citenamefont {Sun}, \citenamefont {Zhao}, \citenamefont {Wang}, \citenamefont {Hong}, \citenamefont {Sidorov}, \citenamefont {Ma}, \citenamefont {Wu}, \citenamefont {Li}, \citenamefont {Meng}, \citenamefont {Sandvik},\ and\ \citenamefont {Sun}}]{Guo_2020}%
  \BibitemOpen
  \bibfield  {author} {\bibinfo {author} {\bibfnamefont {J.}~\bibnamefont {Guo}}, \bibinfo {author} {\bibfnamefont {G.}~\bibnamefont {Sun}}, \bibinfo {author} {\bibfnamefont {B.}~\bibnamefont {Zhao}}, \bibinfo {author} {\bibfnamefont {L.}~\bibnamefont {Wang}}, \bibinfo {author} {\bibfnamefont {W.}~\bibnamefont {Hong}}, \bibinfo {author} {\bibfnamefont {V.~A.}\ \bibnamefont {Sidorov}}, \bibinfo {author} {\bibfnamefont {N.}~\bibnamefont {Ma}}, \bibinfo {author} {\bibfnamefont {Q.}~\bibnamefont {Wu}}, \bibinfo {author} {\bibfnamefont {S.}~\bibnamefont {Li}}, \bibinfo {author} {\bibfnamefont {Z.~Y.}\ \bibnamefont {Meng}}, \bibinfo {author} {\bibfnamefont {A.~W.}\ \bibnamefont {Sandvik}},\ and\ \bibinfo {author} {\bibfnamefont {L.}~\bibnamefont {Sun}},\ }\bibfield  {title} {\bibinfo {title} {Quantum phases of {SrCu$_2$(BO$_3$)$_2$} from high-pressure thermodynamics},\ }\href@noop {} {\bibfield  {journal} {\bibinfo  {journal} {Phys. Rev. Lett.}\ }\textbf {\bibinfo {volume} {124}},\ \bibinfo {pages} {206602}
  (\bibinfo {year} {2020})}\BibitemShut {NoStop}%
\bibitem [{\citenamefont {Sakurai}\ \emph {et~al.}(2009)\citenamefont {Sakurai}, \citenamefont {Tomoo}, \citenamefont {Okubo}, \citenamefont {Ohta}, \citenamefont {Kudo},\ and\ \citenamefont {Koike}}]{Sakurai_2009}%
  \BibitemOpen
  \bibfield  {author} {\bibinfo {author} {\bibfnamefont {T.}~\bibnamefont {Sakurai}}, \bibinfo {author} {\bibfnamefont {M.}~\bibnamefont {Tomoo}}, \bibinfo {author} {\bibfnamefont {S.}~\bibnamefont {Okubo}}, \bibinfo {author} {\bibfnamefont {H.}~\bibnamefont {Ohta}}, \bibinfo {author} {\bibfnamefont {K.}~\bibnamefont {Kudo}},\ and\ \bibinfo {author} {\bibfnamefont {Y.}~\bibnamefont {Koike}},\ }\bibfield  {title} {\bibinfo {title} {High-field and high-pressure esr measurements of {SrCu2(BO3)2}},\ }\href@noop {} {\bibfield  {journal} {\bibinfo  {journal} {Journal of Physics: Conference Series}\ }\textbf {\bibinfo {volume} {150}},\ \bibinfo {pages} {042171} (\bibinfo {year} {2009})}\BibitemShut {NoStop}%
\bibitem [{\citenamefont {Haravifard}\ \emph {et~al.}(2016)\citenamefont {Haravifard}, \citenamefont {Graf}, \citenamefont {Feiguin}, \citenamefont {Batista}, \citenamefont {Lang}, \citenamefont {Silevitch}, \citenamefont {Srajer}, \citenamefont {Gaulin}, \citenamefont {Dabkowska},\ and\ \citenamefont {Rosenbaum}}]{Haravifard_2016}%
  \BibitemOpen
  \bibfield  {author} {\bibinfo {author} {\bibfnamefont {S.}~\bibnamefont {Haravifard}}, \bibinfo {author} {\bibfnamefont {D.}~\bibnamefont {Graf}}, \bibinfo {author} {\bibfnamefont {A.~E.}\ \bibnamefont {Feiguin}}, \bibinfo {author} {\bibfnamefont {C.~D.}\ \bibnamefont {Batista}}, \bibinfo {author} {\bibfnamefont {J.~C.}\ \bibnamefont {Lang}}, \bibinfo {author} {\bibfnamefont {D.~M.}\ \bibnamefont {Silevitch}}, \bibinfo {author} {\bibfnamefont {G.}~\bibnamefont {Srajer}}, \bibinfo {author} {\bibfnamefont {B.~D.}\ \bibnamefont {Gaulin}}, \bibinfo {author} {\bibfnamefont {H.~A.}\ \bibnamefont {Dabkowska}},\ and\ \bibinfo {author} {\bibfnamefont {T.~F.}\ \bibnamefont {Rosenbaum}},\ }\bibfield  {title} {\bibinfo {title} {Crystallization of spin superlattices with pressure and field in the layered magnet {SrCu$_2$(BO$_3$)$_2$}},\ }\href@noop {} {\bibfield  {journal} {\bibinfo  {journal} {Nature Communications}\ }\textbf {\bibinfo {volume} {7}} (\bibinfo {year} {2016})}\BibitemShut {NoStop}%
\bibitem [{\citenamefont {Yang}\ \emph {et~al.}(2022)\citenamefont {Yang}, \citenamefont {Sandvik},\ and\ \citenamefont {Wang}}]{Yang_2022}%
  \BibitemOpen
  \bibfield  {author} {\bibinfo {author} {\bibfnamefont {J.}~\bibnamefont {Yang}}, \bibinfo {author} {\bibfnamefont {A.~W.}\ \bibnamefont {Sandvik}},\ and\ \bibinfo {author} {\bibfnamefont {L.}~\bibnamefont {Wang}},\ }\bibfield  {title} {\bibinfo {title} {Quantum criticality and spin liquid phase in the shastry-sutherland model},\ }\href@noop {} {\bibfield  {journal} {\bibinfo  {journal} {Phys. Rev. B}\ }\textbf {\bibinfo {volume} {105}},\ \bibinfo {pages} {L060409} (\bibinfo {year} {2022})}\BibitemShut {NoStop}%
\bibitem [{\citenamefont {Wang}\ \emph {et~al.}(2022)\citenamefont {Wang}, \citenamefont {Zhang},\ and\ \citenamefont {Sandvik}}]{Wang_2022}%
  \BibitemOpen
  \bibfield  {author} {\bibinfo {author} {\bibfnamefont {L.}~\bibnamefont {Wang}}, \bibinfo {author} {\bibfnamefont {Y.}~\bibnamefont {Zhang}},\ and\ \bibinfo {author} {\bibfnamefont {A.~W.}\ \bibnamefont {Sandvik}},\ }\bibfield  {title} {\bibinfo {title} {Quantum spin liquid phase in the shastry-sutherland model detected by an improved level spectroscopic method},\ }\href@noop {} {\bibfield  {journal} {\bibinfo  {journal} {Chinese Physics Letters}\ }\textbf {\bibinfo {volume} {39}},\ \bibinfo {pages} {077502} (\bibinfo {year} {2022})}\BibitemShut {NoStop}%
\bibitem [{\citenamefont {Miyahara}\ and\ \citenamefont {Ueda}(2000{\natexlab{a}})}]{Miyahara_2000}%
  \BibitemOpen
  \bibfield  {author} {\bibinfo {author} {\bibfnamefont {S.}~\bibnamefont {Miyahara}}\ and\ \bibinfo {author} {\bibfnamefont {K.}~\bibnamefont {Ueda}},\ }\bibfield  {title} {\bibinfo {title} {The magnetization plateaus of {SrCu$_2$(BO$_3$)$_2$}},\ }\href@noop {} {\bibfield  {journal} {\bibinfo  {journal} {Physica B: Condensed Matter}\ }\textbf {\bibinfo {volume} {281-282}},\ \bibinfo {pages} {661} (\bibinfo {year} {2000}{\natexlab{a}})}\BibitemShut {NoStop}%
\bibitem [{\citenamefont {Kageyama}\ \emph {et~al.}(2000)\citenamefont {Kageyama}, \citenamefont {Nishi}, \citenamefont {Aso}, \citenamefont {Onizuka}, \citenamefont {Yosihama}, \citenamefont {Nukui}, \citenamefont {Kodama}, \citenamefont {Kakurai},\ and\ \citenamefont {Ueda}}]{Kageyama_2000}%
  \BibitemOpen
  \bibfield  {author} {\bibinfo {author} {\bibfnamefont {H.}~\bibnamefont {Kageyama}}, \bibinfo {author} {\bibfnamefont {M.}~\bibnamefont {Nishi}}, \bibinfo {author} {\bibfnamefont {N.}~\bibnamefont {Aso}}, \bibinfo {author} {\bibfnamefont {K.}~\bibnamefont {Onizuka}}, \bibinfo {author} {\bibfnamefont {T.}~\bibnamefont {Yosihama}}, \bibinfo {author} {\bibfnamefont {K.}~\bibnamefont {Nukui}}, \bibinfo {author} {\bibfnamefont {K.}~\bibnamefont {Kodama}}, \bibinfo {author} {\bibfnamefont {K.}~\bibnamefont {Kakurai}},\ and\ \bibinfo {author} {\bibfnamefont {Y.}~\bibnamefont {Ueda}},\ }\bibfield  {title} {\bibinfo {title} {Direct evidence for the localized single-triplet excitations and the dispersive multitriplet excitations in {SrCu$_2$(BO$_3$)$_2$}},\ }\href@noop {} {\bibfield  {journal} {\bibinfo  {journal} {Phys. Rev. Lett.}\ }\textbf {\bibinfo {volume} {84}},\ \bibinfo {pages} {5876} (\bibinfo {year} {2000})}\BibitemShut {NoStop}%
\bibitem [{\citenamefont {Miyahara}\ and\ \citenamefont {Ueda}(2000{\natexlab{b}})}]{Miyahara_2000_1}%
  \BibitemOpen
  \bibfield  {author} {\bibinfo {author} {\bibfnamefont {S.}~\bibnamefont {Miyahara}}\ and\ \bibinfo {author} {\bibfnamefont {K.}~\bibnamefont {Ueda}},\ }\bibfield  {title} {\bibinfo {title} {Superstructures at magnetization plateaus in {SrCu$_2$(BO$_3$)$_2$}},\ }\href@noop {} {\bibfield  {journal} {\bibinfo  {journal} {Phys. Rev. B}\ }\textbf {\bibinfo {volume} {61}},\ \bibinfo {pages} {3417} (\bibinfo {year} {2000}{\natexlab{b}})}\BibitemShut {NoStop}%
\bibitem [{\citenamefont {Sebastian}\ \emph {et~al.}(2008)\citenamefont {Sebastian}, \citenamefont {Harrison}, \citenamefont {Sengupta}, \citenamefont {Batista}, \citenamefont {Francoual}, \citenamefont {Palm}, \citenamefont {Murphy}, \citenamefont {Marcano}, \citenamefont {Dabkowska},\ and\ \citenamefont {Gaulin}}]{08_sebastian}%
  \BibitemOpen
  \bibfield  {author} {\bibinfo {author} {\bibfnamefont {S.}~\bibnamefont {Sebastian}}, \bibinfo {author} {\bibfnamefont {N.}~\bibnamefont {Harrison}}, \bibinfo {author} {\bibfnamefont {P.}~\bibnamefont {Sengupta}}, \bibinfo {author} {\bibfnamefont {C.}~\bibnamefont {Batista}}, \bibinfo {author} {\bibfnamefont {S.}~\bibnamefont {Francoual}}, \bibinfo {author} {\bibfnamefont {E.}~\bibnamefont {Palm}}, \bibinfo {author} {\bibfnamefont {T.}~\bibnamefont {Murphy}}, \bibinfo {author} {\bibfnamefont {N.}~\bibnamefont {Marcano}}, \bibinfo {author} {\bibfnamefont {H.}~\bibnamefont {Dabkowska}},\ and\ \bibinfo {author} {\bibfnamefont {B.}~\bibnamefont {Gaulin}},\ }\bibfield  {title} {\bibinfo {title} {Fractalization drives crystalline states in a frustrated spin system},\ }\href@noop {} {\bibfield  {journal} {\bibinfo  {journal} {PNAS}\ }\textbf {\bibinfo {volume} {105}},\ \bibinfo {pages} {20157} (\bibinfo {year} {2008})}\BibitemShut {NoStop}%
\bibitem [{03_()}]{03_ji}%
  \BibitemOpen
  \bibfield  {title} {\bibinfo {title} {Interference of magnetic and anisotropic tensor susceptibility reflections in resonant x-ray scattering of {GdB$_4$}},\ }\href@noop {} {\ }\BibitemShut {NoStop}%
\bibitem [{\citenamefont {Watanuki}\ \emph {et~al.}(2005)\citenamefont {Watanuki}, \citenamefont {Sato}, \citenamefont {Suzuki}, \citenamefont {Ishihara}, \citenamefont {Yanagisawa}, \citenamefont {Nemoto},\ and\ \citenamefont {Goto}}]{05_watanuki}%
  \BibitemOpen
  \bibfield  {author} {\bibinfo {author} {\bibfnamefont {R.}~\bibnamefont {Watanuki}}, \bibinfo {author} {\bibfnamefont {G.}~\bibnamefont {Sato}}, \bibinfo {author} {\bibfnamefont {K.}~\bibnamefont {Suzuki}}, \bibinfo {author} {\bibfnamefont {M.}~\bibnamefont {Ishihara}}, \bibinfo {author} {\bibfnamefont {T.}~\bibnamefont {Yanagisawa}}, \bibinfo {author} {\bibfnamefont {Y.}~\bibnamefont {Nemoto}},\ and\ \bibinfo {author} {\bibfnamefont {T.}~\bibnamefont {Goto}},\ }\bibfield  {title} {\bibinfo {title} {Geometrical quadrupolar frustration in {DyB$_4$}},\ }\href@noop {} {\bibfield  {journal} {\bibinfo  {journal} {Journal of the Physical Society of Japan}\ }\textbf {\bibinfo {volume} {74}},\ \bibinfo {pages} {2169} (\bibinfo {year} {2005})}\BibitemShut {NoStop}%
\bibitem [{\citenamefont {Wigger}\ \emph {et~al.}(2005)\citenamefont {Wigger}, \citenamefont {Felder}, \citenamefont {Monnier}, \citenamefont {Ott}, \citenamefont {Pham},\ and\ \citenamefont {Fisk}}]{05_wigger}%
  \BibitemOpen
  \bibfield  {author} {\bibinfo {author} {\bibfnamefont {G.~A.}\ \bibnamefont {Wigger}}, \bibinfo {author} {\bibfnamefont {E.}~\bibnamefont {Felder}}, \bibinfo {author} {\bibfnamefont {R.}~\bibnamefont {Monnier}}, \bibinfo {author} {\bibfnamefont {H.~R.}\ \bibnamefont {Ott}}, \bibinfo {author} {\bibfnamefont {L.}~\bibnamefont {Pham}},\ and\ \bibinfo {author} {\bibfnamefont {Z.}~\bibnamefont {Fisk}},\ }\bibfield  {title} {\bibinfo {title} {Low-temperature phase transitions in the induced-moment system {PrB$_4$}},\ }\href@noop {} {\bibfield  {journal} {\bibinfo  {journal} {Phys. Rev. B}\ }\textbf {\bibinfo {volume} {72}},\ \bibinfo {pages} {014419} (\bibinfo {year} {2005})}\BibitemShut {NoStop}%
\bibitem [{\citenamefont {Yoshii}\ \emph {et~al.}(2006)\citenamefont {Yoshii}, \citenamefont {Yamamoto}, \citenamefont {Hagiwara}, \citenamefont {Shigekawa}, \citenamefont {Michimura}, \citenamefont {Iga}, \citenamefont {Takabatake},\ and\ \citenamefont {Kindo}}]{06_yoshii}%
  \BibitemOpen
  \bibfield  {author} {\bibinfo {author} {\bibfnamefont {S.}~\bibnamefont {Yoshii}}, \bibinfo {author} {\bibfnamefont {T.}~\bibnamefont {Yamamoto}}, \bibinfo {author} {\bibfnamefont {M.}~\bibnamefont {Hagiwara}}, \bibinfo {author} {\bibfnamefont {A.}~\bibnamefont {Shigekawa}}, \bibinfo {author} {\bibfnamefont {S.}~\bibnamefont {Michimura}}, \bibinfo {author} {\bibfnamefont {F.}~\bibnamefont {Iga}}, \bibinfo {author} {\bibfnamefont {T.}~\bibnamefont {Takabatake}},\ and\ \bibinfo {author} {\bibfnamefont {K.}~\bibnamefont {Kindo}},\ }\bibfield  {title} {\bibinfo {title} {High-field magnetization of {TmB$_4$}},\ }\href {https://dx.doi.org/10.1088/1742-6596/51/1/011} {\bibfield  {journal} {\bibinfo  {journal} {Journal of Physics: Conference Series}\ }\textbf {\bibinfo {volume} {51}},\ \bibinfo {pages} {59} (\bibinfo {year} {2006})}\BibitemShut {NoStop}%
\bibitem [{\citenamefont {Michimura}\ \emph {et~al.}(2006)\citenamefont {Michimura}, \citenamefont {Shigekawa}, \citenamefont {Iga}, \citenamefont {Sera}, \citenamefont {Takabatake}, \citenamefont {Ohoyama},\ and\ \citenamefont {Okabe}}]{06_michimura}%
  \BibitemOpen
  \bibfield  {author} {\bibinfo {author} {\bibfnamefont {S.}~\bibnamefont {Michimura}}, \bibinfo {author} {\bibfnamefont {A.}~\bibnamefont {Shigekawa}}, \bibinfo {author} {\bibfnamefont {F.}~\bibnamefont {Iga}}, \bibinfo {author} {\bibfnamefont {M.}~\bibnamefont {Sera}}, \bibinfo {author} {\bibfnamefont {T.}~\bibnamefont {Takabatake}}, \bibinfo {author} {\bibfnamefont {K.}~\bibnamefont {Ohoyama}},\ and\ \bibinfo {author} {\bibfnamefont {Y.}~\bibnamefont {Okabe}},\ }\bibfield  {title} {\bibinfo {title} {Magnetic frustrations in the shastry–sutherland system {ErB$_4$}},\ }\href@noop {} {\bibfield  {journal} {\bibinfo  {journal} {Physica B: Condensed Matter}\ }\textbf {\bibinfo {volume} {378-380}},\ \bibinfo {pages} {596} (\bibinfo {year} {2006})},\ \bibinfo {note} {proceedings of the International Conference on Strongly Correlated Electron Systems}\BibitemShut {NoStop}%
\bibitem [{\citenamefont {Matsumura}\ \emph {et~al.}(2007)\citenamefont {Matsumura}, \citenamefont {Okuyama},\ and\ \citenamefont {Murakami}}]{07_matsumura}%
  \BibitemOpen
  \bibfield  {author} {\bibinfo {author} {\bibfnamefont {T.}~\bibnamefont {Matsumura}}, \bibinfo {author} {\bibfnamefont {D.}~\bibnamefont {Okuyama}},\ and\ \bibinfo {author} {\bibfnamefont {Y.}~\bibnamefont {Murakami}},\ }\bibfield  {title} {\bibinfo {title} {Non-collinear magnetic structure of tbb4},\ }\href@noop {} {\bibfield  {journal} {\bibinfo  {journal} {Journal of the Physical Society of Japan}\ }\textbf {\bibinfo {volume} {76}},\ \bibinfo {pages} {015001} (\bibinfo {year} {2007})}\BibitemShut {NoStop}%
\bibitem [{\citenamefont {Okuyama}\ \emph {et~al.}(2007)\citenamefont {Okuyama}, \citenamefont {Matsumura}, \citenamefont {Iwasa},\ and\ \citenamefont {Murakami}}]{07_okuyama}%
  \BibitemOpen
  \bibfield  {author} {\bibinfo {author} {\bibfnamefont {D.}~\bibnamefont {Okuyama}}, \bibinfo {author} {\bibfnamefont {T.}~\bibnamefont {Matsumura}}, \bibinfo {author} {\bibfnamefont {K.}~\bibnamefont {Iwasa}},\ and\ \bibinfo {author} {\bibfnamefont {Y.}~\bibnamefont {Murakami}},\ }\bibfield  {title} {\bibinfo {title} {Magnetic phase transition in {HoB$_4$} studied by neutron diffraction},\ }\href@noop {} {\bibfield  {journal} {\bibinfo  {journal} {Journal of Magnetism and Magnetic Materials}\ }\textbf {\bibinfo {volume} {310}},\ \bibinfo {pages} {e152} (\bibinfo {year} {2007})}\BibitemShut {NoStop}%
\bibitem [{\citenamefont {Kim}\ \emph {et~al.}(2010)\citenamefont {Kim}, \citenamefont {Sung}, \citenamefont {Kang}, \citenamefont {Kim}, \citenamefont {Cho},\ and\ \citenamefont {Rhyee}}]{10_kim}%
  \BibitemOpen
  \bibfield  {author} {\bibinfo {author} {\bibfnamefont {J.~Y.}\ \bibnamefont {Kim}}, \bibinfo {author} {\bibfnamefont {N.~H.}\ \bibnamefont {Sung}}, \bibinfo {author} {\bibfnamefont {B.~Y.}\ \bibnamefont {Kang}}, \bibinfo {author} {\bibfnamefont {M.~S.}\ \bibnamefont {Kim}}, \bibinfo {author} {\bibfnamefont {B.~K.}\ \bibnamefont {Cho}},\ and\ \bibinfo {author} {\bibfnamefont {J.-S.}\ \bibnamefont {Rhyee}},\ }\bibfield  {title} {\bibinfo {title} {{Magnetic anisotropy and magnon gap state of SmB4 single crystal}},\ }\href@noop {} {\bibfield  {journal} {\bibinfo  {journal} {Journal of Applied Physics}\ }\textbf {\bibinfo {volume} {107}},\ \bibinfo {pages} {09E111} (\bibinfo {year} {2010})}\BibitemShut {NoStop}%
\bibitem [{\citenamefont {Yamauchi}\ \emph {et~al.}(2017)\citenamefont {Yamauchi}, \citenamefont {Metoki}, \citenamefont {Watanuki}, \citenamefont {Suzuki}, \citenamefont {Fukazawa}, \citenamefont {Chi},\ and\ \citenamefont {Fernandez-Baca}}]{17_yamauchi}%
  \BibitemOpen
  \bibfield  {author} {\bibinfo {author} {\bibfnamefont {H.}~\bibnamefont {Yamauchi}}, \bibinfo {author} {\bibfnamefont {N.}~\bibnamefont {Metoki}}, \bibinfo {author} {\bibfnamefont {R.}~\bibnamefont {Watanuki}}, \bibinfo {author} {\bibfnamefont {K.}~\bibnamefont {Suzuki}}, \bibinfo {author} {\bibfnamefont {H.}~\bibnamefont {Fukazawa}}, \bibinfo {author} {\bibfnamefont {S.}~\bibnamefont {Chi}},\ and\ \bibinfo {author} {\bibfnamefont {J.~A.}\ \bibnamefont {Fernandez-Baca}},\ }\bibfield  {title} {\bibinfo {title} {Magnetic structure and quadrupolar order parameter driven by geometrical frustration effect in {NdB$_4$}},\ }\href@noop {} {\bibfield  {journal} {\bibinfo  {journal} {Journal of the Physical Society of Japan}\ }\textbf {\bibinfo {volume} {86}},\ \bibinfo {pages} {044705} (\bibinfo {year} {2017})}\BibitemShut {NoStop}%
\bibitem [{\citenamefont {Fisk}\ \emph {et~al.}(1981)\citenamefont {Fisk}, \citenamefont {Maple}, \citenamefont {Johnston},\ and\ \citenamefont {Woolf}}]{81_fisk}%
  \BibitemOpen
  \bibfield  {author} {\bibinfo {author} {\bibfnamefont {Z.}~\bibnamefont {Fisk}}, \bibinfo {author} {\bibfnamefont {M.}~\bibnamefont {Maple}}, \bibinfo {author} {\bibfnamefont {D.}~\bibnamefont {Johnston}},\ and\ \bibinfo {author} {\bibfnamefont {L.}~\bibnamefont {Woolf}},\ }\bibfield  {title} {\bibinfo {title} {Multiple phase transitions in rare earth tetraborides at low temperature},\ }\href@noop {} {\bibfield  {journal} {\bibinfo  {journal} {Solid State Communications}\ }\textbf {\bibinfo {volume} {39}},\ \bibinfo {pages} {1189} (\bibinfo {year} {1981})}\BibitemShut {NoStop}%
\bibitem [{\citenamefont {Gianduzzo}\ \emph {et~al.}(1981)\citenamefont {Gianduzzo}, \citenamefont {Georges}, \citenamefont {Chevalier}, \citenamefont {Etourneau}, \citenamefont {Hagenmuller}, \citenamefont {Will},\ and\ \citenamefont {Schäfer}}]{81_gianduzzo}%
  \BibitemOpen
  \bibfield  {author} {\bibinfo {author} {\bibfnamefont {J.}~\bibnamefont {Gianduzzo}}, \bibinfo {author} {\bibfnamefont {R.}~\bibnamefont {Georges}}, \bibinfo {author} {\bibfnamefont {B.}~\bibnamefont {Chevalier}}, \bibinfo {author} {\bibfnamefont {J.}~\bibnamefont {Etourneau}}, \bibinfo {author} {\bibfnamefont {P.}~\bibnamefont {Hagenmuller}}, \bibinfo {author} {\bibfnamefont {G.}~\bibnamefont {Will}},\ and\ \bibinfo {author} {\bibfnamefont {W.}~\bibnamefont {Schäfer}},\ }\bibfield  {title} {\bibinfo {title} {Anisotropy and magnetic phase transitions in the rare earth tetraborides {TbB$_4$}, {HoB$_4$} and {ErB$_4$}},\ }\href@noop {} {\bibfield  {journal} {\bibinfo  {journal} {Journal of the Less Common Metals}\ }\textbf {\bibinfo {volume} {82}},\ \bibinfo {pages} {29} (\bibinfo {year} {1981})}\BibitemShut {NoStop}%
\bibitem [{\citenamefont {Will}\ \emph {et~al.}(1981)\citenamefont {Will}, \citenamefont {Schäfer}, \citenamefont {Pfeiffer}, \citenamefont {Elf},\ and\ \citenamefont {Etourneau}}]{81_will}%
  \BibitemOpen
  \bibfield  {author} {\bibinfo {author} {\bibfnamefont {G.}~\bibnamefont {Will}}, \bibinfo {author} {\bibfnamefont {W.}~\bibnamefont {Schäfer}}, \bibinfo {author} {\bibfnamefont {F.}~\bibnamefont {Pfeiffer}}, \bibinfo {author} {\bibfnamefont {F.}~\bibnamefont {Elf}},\ and\ \bibinfo {author} {\bibfnamefont {J.}~\bibnamefont {Etourneau}},\ }\bibfield  {title} {\bibinfo {title} {Neutron diffraction studies of {TbB$_4$} and {ErB$_4$}},\ }\href@noop {} {\bibfield  {journal} {\bibinfo  {journal} {Journal of the Less Common Metals}\ }\textbf {\bibinfo {volume} {82}},\ \bibinfo {pages} {349} (\bibinfo {year} {1981})}\BibitemShut {NoStop}%
\bibitem [{\citenamefont {Iga}\ \emph {et~al.}(2007)\citenamefont {Iga}, \citenamefont {Shigekawa}, \citenamefont {Hasegawa}, \citenamefont {Michimura}, \citenamefont {Takabatake}, \citenamefont {Yoshii}, \citenamefont {Yamamoto}, \citenamefont {Hagiwara},\ and\ \citenamefont {Kindo}}]{07_iga}%
  \BibitemOpen
  \bibfield  {author} {\bibinfo {author} {\bibfnamefont {F.}~\bibnamefont {Iga}}, \bibinfo {author} {\bibfnamefont {A.}~\bibnamefont {Shigekawa}}, \bibinfo {author} {\bibfnamefont {Y.}~\bibnamefont {Hasegawa}}, \bibinfo {author} {\bibfnamefont {S.}~\bibnamefont {Michimura}}, \bibinfo {author} {\bibfnamefont {T.}~\bibnamefont {Takabatake}}, \bibinfo {author} {\bibfnamefont {S.}~\bibnamefont {Yoshii}}, \bibinfo {author} {\bibfnamefont {T.}~\bibnamefont {Yamamoto}}, \bibinfo {author} {\bibfnamefont {M.}~\bibnamefont {Hagiwara}},\ and\ \bibinfo {author} {\bibfnamefont {K.}~\bibnamefont {Kindo}},\ }\bibfield  {title} {\bibinfo {title} {Highly anisotropic magnetic phase diagram of a 2-dimensional orthogonal dimer system {TmB$_4$}},\ }\href@noop {} {\bibfield  {journal} {\bibinfo  {journal} {Journal of Magnetism and Magnetic Materials}\ }\textbf {\bibinfo {volume} {310}},\ \bibinfo {pages} {e443} (\bibinfo {year} {2007})},\ \bibinfo {note} {proceedings of the 17th International Conference on Magnetism}\BibitemShut
  {NoStop}%
\bibitem [{\citenamefont {Okuyama}\ \emph {et~al.}(2008)\citenamefont {Okuyama}, \citenamefont {Matsumura}, \citenamefont {Mouri}, \citenamefont {Ishikawa}, \citenamefont {Ohoyama}, \citenamefont {Hiraka}, \citenamefont {Nakao}, \citenamefont {Iwasa},\ and\ \citenamefont {Murakami}}]{08_okuyama}%
  \BibitemOpen
  \bibfield  {author} {\bibinfo {author} {\bibfnamefont {D.}~\bibnamefont {Okuyama}}, \bibinfo {author} {\bibfnamefont {T.}~\bibnamefont {Matsumura}}, \bibinfo {author} {\bibfnamefont {T.}~\bibnamefont {Mouri}}, \bibinfo {author} {\bibfnamefont {N.}~\bibnamefont {Ishikawa}}, \bibinfo {author} {\bibfnamefont {K.}~\bibnamefont {Ohoyama}}, \bibinfo {author} {\bibfnamefont {H.}~\bibnamefont {Hiraka}}, \bibinfo {author} {\bibfnamefont {H.}~\bibnamefont {Nakao}}, \bibinfo {author} {\bibfnamefont {K.}~\bibnamefont {Iwasa}},\ and\ \bibinfo {author} {\bibfnamefont {Y.}~\bibnamefont {Murakami}},\ }\bibfield  {title} {\bibinfo {title} {Competition of magnetic and quadrupolar order parameters in {HoB$_4$}},\ }\href@noop {} {\bibfield  {journal} {\bibinfo  {journal} {Journal of the Physical Society of Japan}\ }\textbf {\bibinfo {volume} {77}},\ \bibinfo {pages} {044709} (\bibinfo {year} {2008})}\BibitemShut {NoStop}%
\bibitem [{\citenamefont {Sim}\ \emph {et~al.}(2016)\citenamefont {Sim}, \citenamefont {Lee}, \citenamefont {Hong}, \citenamefont {Jeong}, \citenamefont {Zhang}, \citenamefont {Kamiyama}, \citenamefont {Adroja}, \citenamefont {Murray}, \citenamefont {Thompson}, \citenamefont {Iga}, \citenamefont {Ji}, \citenamefont {Khomskii},\ and\ \citenamefont {Park}}]{16_sim}%
  \BibitemOpen
  \bibfield  {author} {\bibinfo {author} {\bibfnamefont {H.}~\bibnamefont {Sim}}, \bibinfo {author} {\bibfnamefont {S.}~\bibnamefont {Lee}}, \bibinfo {author} {\bibfnamefont {K.-P.}\ \bibnamefont {Hong}}, \bibinfo {author} {\bibfnamefont {J.}~\bibnamefont {Jeong}}, \bibinfo {author} {\bibfnamefont {J.~R.}\ \bibnamefont {Zhang}}, \bibinfo {author} {\bibfnamefont {T.}~\bibnamefont {Kamiyama}}, \bibinfo {author} {\bibfnamefont {D.~T.}\ \bibnamefont {Adroja}}, \bibinfo {author} {\bibfnamefont {C.~A.}\ \bibnamefont {Murray}}, \bibinfo {author} {\bibfnamefont {S.~P.}\ \bibnamefont {Thompson}}, \bibinfo {author} {\bibfnamefont {F.}~\bibnamefont {Iga}}, \bibinfo {author} {\bibfnamefont {S.}~\bibnamefont {Ji}}, \bibinfo {author} {\bibfnamefont {D.}~\bibnamefont {Khomskii}},\ and\ \bibinfo {author} {\bibfnamefont {J.-G.}\ \bibnamefont {Park}},\ }\bibfield  {title} {\bibinfo {title} {Spontaneous structural distortion of the metallic shastry-sutherland system {DyB$_4$} by quadrupole-spin-lattice coupling},\ }\href@noop
  {} {\bibfield  {journal} {\bibinfo  {journal} {Phys. Rev. B}\ }\textbf {\bibinfo {volume} {94}},\ \bibinfo {pages} {195128} (\bibinfo {year} {2016})}\BibitemShut {NoStop}%
\bibitem [{\citenamefont {Blanco}\ \emph {et~al.}(2006)\citenamefont {Blanco}, \citenamefont {Brown}, \citenamefont {Stunault}, \citenamefont {Katsumata}, \citenamefont {Iga},\ and\ \citenamefont {Michimura}}]{06_blanco}%
  \BibitemOpen
  \bibfield  {author} {\bibinfo {author} {\bibfnamefont {J.~A.}\ \bibnamefont {Blanco}}, \bibinfo {author} {\bibfnamefont {P.}~\bibnamefont {Brown}}, \bibinfo {author} {\bibfnamefont {A.}~\bibnamefont {Stunault}}, \bibinfo {author} {\bibfnamefont {K.}~\bibnamefont {Katsumata}}, \bibinfo {author} {\bibfnamefont {F.}~\bibnamefont {Iga}},\ and\ \bibinfo {author} {\bibfnamefont {S.}~\bibnamefont {Michimura}},\ }\bibfield  {title} {\bibinfo {title} {Magnetic structure of {GdB$_4$} from spherical neutron polarimetry},\ }\href@noop {} {\bibfield  {journal} {\bibinfo  {journal} {Phys. Rev. B}\ }\textbf {\bibinfo {volume} {73}},\ \bibinfo {pages} {212411} (\bibinfo {year} {2006})}\BibitemShut {NoStop}%
\bibitem [{\citenamefont {Hulliger}(1995)}]{95_hulliger}%
  \BibitemOpen
  \bibfield  {author} {\bibinfo {author} {\bibfnamefont {F.}~\bibnamefont {Hulliger}},\ }\bibfield  {title} {\bibinfo {title} {On new {Mo$_2$FeB$_2$}-type representatives {Ln$_2$Rh$_2$In}},\ }\href@noop {} {\bibfield  {journal} {\bibinfo  {journal} {Journal of Alloys and Compounds}\ }\textbf {\bibinfo {volume} {221}},\ \bibinfo {pages} {L11} (\bibinfo {year} {1995})}\BibitemShut {NoStop}%
\bibitem [{\citenamefont {Giovannini}\ \emph {et~al.}(1998)\citenamefont {Giovannini}, \citenamefont {Michor}, \citenamefont {Bauer}, \citenamefont {Hilscher}, \citenamefont {Rogl},\ and\ \citenamefont {Ferro}}]{98_giovannini}%
  \BibitemOpen
  \bibfield  {author} {\bibinfo {author} {\bibfnamefont {M.}~\bibnamefont {Giovannini}}, \bibinfo {author} {\bibfnamefont {H.}~\bibnamefont {Michor}}, \bibinfo {author} {\bibfnamefont {E.}~\bibnamefont {Bauer}}, \bibinfo {author} {\bibfnamefont {G.}~\bibnamefont {Hilscher}}, \bibinfo {author} {\bibfnamefont {P.}~\bibnamefont {Rogl}},\ and\ \bibinfo {author} {\bibfnamefont {R.}~\bibnamefont {Ferro}},\ }\bibfield  {title} {\bibinfo {title} {Structural chemistry, magnetism and thermodynamic properties of {R$_2$Pd$_2$In}},\ }\href@noop {} {\bibfield  {journal} {\bibinfo  {journal} {Journal of Alloys and Compounds}\ }\textbf {\bibinfo {volume} {280}},\ \bibinfo {pages} {26} (\bibinfo {year} {1998})}\BibitemShut {NoStop}%
\bibitem [{\citenamefont {Kraft}\ \emph {et~al.}(2003)\citenamefont {Kraft}, \citenamefont {Fickenscher}, \citenamefont {Kotzyba}, \citenamefont {Hoffmann},\ and\ \citenamefont {Pöttgen}}]{03_kraft}%
  \BibitemOpen
  \bibfield  {author} {\bibinfo {author} {\bibfnamefont {R.}~\bibnamefont {Kraft}}, \bibinfo {author} {\bibfnamefont {T.}~\bibnamefont {Fickenscher}}, \bibinfo {author} {\bibfnamefont {G.}~\bibnamefont {Kotzyba}}, \bibinfo {author} {\bibfnamefont {R.-D.}\ \bibnamefont {Hoffmann}},\ and\ \bibinfo {author} {\bibfnamefont {R.}~\bibnamefont {Pöttgen}},\ }\bibfield  {title} {\bibinfo {title} {Intermetallic rare earth (re) magnesium compounds {REPdMg} and {Re$_2$Pd$_2$Mg}},\ }\href@noop {} {\bibfield  {journal} {\bibinfo  {journal} {Intermetallics}\ }\textbf {\bibinfo {volume} {11}},\ \bibinfo {pages} {111} (\bibinfo {year} {2003})}\BibitemShut {NoStop}%
\bibitem [{\citenamefont {Zaremba}\ \emph {et~al.}(2004)\citenamefont {Zaremba}, \citenamefont {Kaczorowski}, \citenamefont {Nychyporuk}, \citenamefont {Rodewald},\ and\ \citenamefont {Pöttgen}}]{04_zaremba}%
  \BibitemOpen
  \bibfield  {author} {\bibinfo {author} {\bibfnamefont {V.~I.}\ \bibnamefont {Zaremba}}, \bibinfo {author} {\bibfnamefont {D.}~\bibnamefont {Kaczorowski}}, \bibinfo {author} {\bibfnamefont {G.~P.}\ \bibnamefont {Nychyporuk}}, \bibinfo {author} {\bibfnamefont {U.~C.}\ \bibnamefont {Rodewald}},\ and\ \bibinfo {author} {\bibfnamefont {R.}~\bibnamefont {Pöttgen}},\ }\bibfield  {title} {\bibinfo {title} {Structure and physical properties of {Re$_2$Ge$_2$In} {(RE = La, Ce, Pr, Nd)}},\ }\href@noop {} {\bibfield  {journal} {\bibinfo  {journal} {Solid State Sciences}\ }\textbf {\bibinfo {volume} {6}},\ \bibinfo {pages} {1301} (\bibinfo {year} {2004})}\BibitemShut {NoStop}%
\bibitem [{\citenamefont {Rayaprol}\ \emph {et~al.}(2006)\citenamefont {Rayaprol}, \citenamefont {Doğan},\ and\ \citenamefont {Pöttgen}}]{06_rayapol}%
  \BibitemOpen
  \bibfield  {author} {\bibinfo {author} {\bibfnamefont {S.}~\bibnamefont {Rayaprol}}, \bibinfo {author} {\bibfnamefont {A.}~\bibnamefont {Doğan}},\ and\ \bibinfo {author} {\bibfnamefont {R.}~\bibnamefont {Pöttgen}},\ }\bibfield  {title} {\bibinfo {title} {Magnetic properties and specific heat studies of {Re$_2$Pd$_2$Cd} {(RE = La,Ce,Nd)}},\ }\href@noop {} {\bibfield  {journal} {\bibinfo  {journal} {Journal of Physics: Condensed Matter}\ }\textbf {\bibinfo {volume} {18}},\ \bibinfo {pages} {5473} (\bibinfo {year} {2006})}\BibitemShut {NoStop}%
\bibitem [{\citenamefont {Kumar}\ \emph {et~al.}(2008)\citenamefont {Kumar}, \citenamefont {Singh}, \citenamefont {Suresh},\ and\ \citenamefont {Nigam}}]{08_kumar}%
  \BibitemOpen
  \bibfield  {author} {\bibinfo {author} {\bibfnamefont {P.}~\bibnamefont {Kumar}}, \bibinfo {author} {\bibfnamefont {N.~K.}\ \bibnamefont {Singh}}, \bibinfo {author} {\bibfnamefont {K.~G.}\ \bibnamefont {Suresh}},\ and\ \bibinfo {author} {\bibfnamefont {A.~K.}\ \bibnamefont {Nigam}},\ }\bibfield  {title} {\bibinfo {title} {Magnetocaloric and magnetotransport properties of {Re$_2$Ni$_2$Sn} compounds {(Re= Nd, Sm, Gd, and Tb)}},\ }\href@noop {} {\bibfield  {journal} {\bibinfo  {journal} {Phys. Rev. B}\ }\textbf {\bibinfo {volume} {77}},\ \bibinfo {pages} {184411} (\bibinfo {year} {2008})}\BibitemShut {NoStop}%
\bibitem [{\citenamefont {Schappacher}\ \emph {et~al.}(2009)\citenamefont {Schappacher}, \citenamefont {Hermes},\ and\ \citenamefont {Pöttgen}}]{09_schappacher}%
  \BibitemOpen
  \bibfield  {author} {\bibinfo {author} {\bibfnamefont {F.~M.}\ \bibnamefont {Schappacher}}, \bibinfo {author} {\bibfnamefont {W.}~\bibnamefont {Hermes}},\ and\ \bibinfo {author} {\bibfnamefont {R.}~\bibnamefont {Pöttgen}},\ }\bibfield  {title} {\bibinfo {title} {Structure and magnetic properties of {Re$_2$Cu$_2$Cd}},\ }\href@noop {} {\bibfield  {journal} {\bibinfo  {journal} {Journal of Solid State Chemistry}\ }\textbf {\bibinfo {volume} {182}},\ \bibinfo {pages} {265} (\bibinfo {year} {2009})}\BibitemShut {NoStop}%
\bibitem [{\citenamefont {Shah}\ \emph {et~al.}(2009)\citenamefont {Shah}, \citenamefont {Bonville}, \citenamefont {Manfrinetti}, \citenamefont {Wrubl},\ and\ \citenamefont {Dhar}}]{09_shah}%
  \BibitemOpen
  \bibfield  {author} {\bibinfo {author} {\bibfnamefont {K.~V.}\ \bibnamefont {Shah}}, \bibinfo {author} {\bibfnamefont {P.}~\bibnamefont {Bonville}}, \bibinfo {author} {\bibfnamefont {P.}~\bibnamefont {Manfrinetti}}, \bibinfo {author} {\bibfnamefont {F.}~\bibnamefont {Wrubl}},\ and\ \bibinfo {author} {\bibfnamefont {S.~K.}\ \bibnamefont {Dhar}},\ }\bibfield  {title} {\bibinfo {title} {The {Yb$_2$Al$_{1−x}$Mg$_x$Si$_2$} series from a spin fluctuation (x = 0) to a magnetically ordered ground state (x = 1)},\ }\href@noop {} {\bibfield  {journal} {\bibinfo  {journal} {Journal of Physics: Condensed Matter}\ }\textbf {\bibinfo {volume} {21}},\ \bibinfo {pages} {176001} (\bibinfo {year} {2009})}\BibitemShut {NoStop}%
\bibitem [{\citenamefont {Suen}\ \emph {et~al.}(2011)\citenamefont {Suen}, \citenamefont {Tobash},\ and\ \citenamefont {Bobev}}]{11_suen}%
  \BibitemOpen
  \bibfield  {author} {\bibinfo {author} {\bibfnamefont {N.-T.}\ \bibnamefont {Suen}}, \bibinfo {author} {\bibfnamefont {P.~H.}\ \bibnamefont {Tobash}},\ and\ \bibinfo {author} {\bibfnamefont {S.}~\bibnamefont {Bobev}},\ }\bibfield  {title} {\bibinfo {title} {Synthesis, structural characterization and magnetic properties of {Re$_2$MgGe$_2$} {(Re=rare-earth metal)}},\ }\href@noop {} {\bibfield  {journal} {\bibinfo  {journal} {Journal of Solid State Chemistry}\ }\textbf {\bibinfo {volume} {184}},\ \bibinfo {pages} {2941} (\bibinfo {year} {2011})}\BibitemShut {NoStop}%
\bibitem [{\citenamefont {Shimura}\ \emph {et~al.}(2012)\citenamefont {Shimura}, \citenamefont {Sakakibara}, \citenamefont {Iwakawa}, \citenamefont {Sugiyama},\ and\ \citenamefont {Ōnuki}}]{12_shimura}%
  \BibitemOpen
  \bibfield  {author} {\bibinfo {author} {\bibfnamefont {Y.}~\bibnamefont {Shimura}}, \bibinfo {author} {\bibfnamefont {T.}~\bibnamefont {Sakakibara}}, \bibinfo {author} {\bibfnamefont {K.}~\bibnamefont {Iwakawa}}, \bibinfo {author} {\bibfnamefont {K.}~\bibnamefont {Sugiyama}},\ and\ \bibinfo {author} {\bibfnamefont {Y.}~\bibnamefont {Ōnuki}},\ }\bibfield  {title} {\bibinfo {title} {Low temperature magnetization of {Yb$_2$Pt$_2$Pb} with the shastry–sutherland type lattice and a high-rank multipole interaction},\ }\href@noop {} {\bibfield  {journal} {\bibinfo  {journal} {Journal of the Physical Society of Japan}\ }\textbf {\bibinfo {volume} {81}},\ \bibinfo {pages} {103601} (\bibinfo {year} {2012})}\BibitemShut {NoStop}%
\bibitem [{\citenamefont {Miiller}\ \emph {et~al.}(2016)\citenamefont {Miiller}, \citenamefont {Wu}, \citenamefont {Kim}, \citenamefont {Lynn}, \citenamefont {Zaliznyak},\ and\ \citenamefont {Aronson}}]{16_miiller}%
  \BibitemOpen
  \bibfield  {author} {\bibinfo {author} {\bibfnamefont {W.}~\bibnamefont {Miiller}}, \bibinfo {author} {\bibfnamefont {L.~S.}\ \bibnamefont {Wu}}, \bibinfo {author} {\bibnamefont {Kim}}, \bibinfo {author} {\bibfnamefont {J.~W.}\ \bibnamefont {Lynn}}, \bibinfo {author} {\bibfnamefont {I.}~\bibnamefont {Zaliznyak}},\ and\ \bibinfo {author} {\bibfnamefont {M.~C.}\ \bibnamefont {Aronson}},\ }\bibfield  {title} {\bibinfo {title} {Magnetic structure of {Yb$_2$Pt$_2$Pb}. ising moments on the shastry-sutherland lattice},\ }\href@noop {} {\bibfield  {journal} {\bibinfo  {journal} {Phys. Rev. B}\ }\textbf {\bibinfo {volume} {93}},\ \bibinfo {pages} {104419} (\bibinfo {year} {2016})}\BibitemShut {NoStop}%
\bibitem [{\citenamefont {Wu}\ \emph {et~al.}(2016)\citenamefont {Wu}, \citenamefont {Gannon}, \citenamefont {Zaliznyak}, \citenamefont {Tsvelik}, \citenamefont {Brockmann}, \citenamefont {Caux}, \citenamefont {Kim}, \citenamefont {Qiu}, \citenamefont {Copley}, \citenamefont {Ehlers}, \citenamefont {Podlesnyak},\ and\ \citenamefont {Aronson}}]{16_wu}%
  \BibitemOpen
  \bibfield  {author} {\bibinfo {author} {\bibfnamefont {L.~S.}\ \bibnamefont {Wu}}, \bibinfo {author} {\bibfnamefont {W.~J.}\ \bibnamefont {Gannon}}, \bibinfo {author} {\bibfnamefont {I.~A.}\ \bibnamefont {Zaliznyak}}, \bibinfo {author} {\bibfnamefont {A.~M.}\ \bibnamefont {Tsvelik}}, \bibinfo {author} {\bibfnamefont {M.}~\bibnamefont {Brockmann}}, \bibinfo {author} {\bibfnamefont {J.-S.}\ \bibnamefont {Caux}}, \bibinfo {author} {\bibfnamefont {M.~S.}\ \bibnamefont {Kim}}, \bibinfo {author} {\bibfnamefont {Y.}~\bibnamefont {Qiu}}, \bibinfo {author} {\bibfnamefont {J.~R.~D.}\ \bibnamefont {Copley}}, \bibinfo {author} {\bibfnamefont {G.}~\bibnamefont {Ehlers}}, \bibinfo {author} {\bibfnamefont {A.}~\bibnamefont {Podlesnyak}},\ and\ \bibinfo {author} {\bibfnamefont {M.~C.}\ \bibnamefont {Aronson}},\ }\bibfield  {title} {\bibinfo {title} {Orbital-exchange and fractional quantum number excitations in an f-electron metal{Yb$_2$Pt$_2$Pb}},\ }\href@noop {} {\bibfield  {journal} {\bibinfo  {journal} {Science}\
  }\textbf {\bibinfo {volume} {352}},\ \bibinfo {pages} {1206} (\bibinfo {year} {2016})}\BibitemShut {NoStop}%
\bibitem [{\citenamefont {Gannon}\ \emph {et~al.}(2018)\citenamefont {Gannon}, \citenamefont {Chen}, \citenamefont {Sundermann}, \citenamefont {Strigari}, \citenamefont {Utsumi}, \citenamefont {Tsuei}, \citenamefont {Rueff}, \citenamefont {Bencok}, \citenamefont {Tanaka}, \citenamefont {Severing},\ and\ \citenamefont {Aronson}}]{18_gannon}%
  \BibitemOpen
  \bibfield  {author} {\bibinfo {author} {\bibfnamefont {W.~J.}\ \bibnamefont {Gannon}}, \bibinfo {author} {\bibfnamefont {K.}~\bibnamefont {Chen}}, \bibinfo {author} {\bibfnamefont {M.}~\bibnamefont {Sundermann}}, \bibinfo {author} {\bibfnamefont {F.}~\bibnamefont {Strigari}}, \bibinfo {author} {\bibfnamefont {Y.}~\bibnamefont {Utsumi}}, \bibinfo {author} {\bibfnamefont {K.-D.}\ \bibnamefont {Tsuei}}, \bibinfo {author} {\bibfnamefont {J.-P.}\ \bibnamefont {Rueff}}, \bibinfo {author} {\bibfnamefont {P.}~\bibnamefont {Bencok}}, \bibinfo {author} {\bibfnamefont {A.}~\bibnamefont {Tanaka}}, \bibinfo {author} {\bibfnamefont {A.}~\bibnamefont {Severing}},\ and\ \bibinfo {author} {\bibfnamefont {M.~C.}\ \bibnamefont {Aronson}},\ }\bibfield  {title} {\bibinfo {title} {Intermediate valence in single crystalline {Yb$_2$Si$_2$Al}},\ }\href@noop {} {\bibfield  {journal} {\bibinfo  {journal} {Phys. Rev. B}\ }\textbf {\bibinfo {volume} {98}},\ \bibinfo {pages} {075101} (\bibinfo {year} {2018})}\BibitemShut {NoStop}%
\bibitem [{\citenamefont {Gannon}\ \emph {et~al.}(2019{\natexlab{a}})\citenamefont {Gannon}, \citenamefont {Zaliznyak}, \citenamefont {Wu}, \citenamefont {Feiguin}, \citenamefont {Tsvelik}, \citenamefont {Demmel}, \citenamefont {Qiu}, \citenamefont {Copley}, \citenamefont {Kim},\ and\ \citenamefont {Aronson}}]{19_gannon}%
  \BibitemOpen
  \bibfield  {author} {\bibinfo {author} {\bibfnamefont {W.~J.}\ \bibnamefont {Gannon}}, \bibinfo {author} {\bibfnamefont {I.~A.}\ \bibnamefont {Zaliznyak}}, \bibinfo {author} {\bibfnamefont {L.~S.}\ \bibnamefont {Wu}}, \bibinfo {author} {\bibfnamefont {A.~E.}\ \bibnamefont {Feiguin}}, \bibinfo {author} {\bibfnamefont {A.~M.}\ \bibnamefont {Tsvelik}}, \bibinfo {author} {\bibfnamefont {F.}~\bibnamefont {Demmel}}, \bibinfo {author} {\bibfnamefont {Y.}~\bibnamefont {Qiu}}, \bibinfo {author} {\bibfnamefont {J.~R.~D.}\ \bibnamefont {Copley}}, \bibinfo {author} {\bibfnamefont {M.~S.}\ \bibnamefont {Kim}},\ and\ \bibinfo {author} {\bibfnamefont {M.~C.}\ \bibnamefont {Aronson}},\ }\bibfield  {title} {\bibinfo {title} {Spinon confinement and a sharp longitudinal mode in {Yb$_2$Pt$_2$Pb} in magnetic fields},\ }\href@noop {} {\bibfield  {journal} {\bibinfo  {journal} {Nature Communications}\ }\textbf {\bibinfo {volume} {10}},\ \bibinfo {pages} {1123} (\bibinfo {year} {2019}{\natexlab{a}})}\BibitemShut {NoStop}%
\bibitem [{\citenamefont {Wakeshima}\ \emph {et~al.}(2003)\citenamefont {Wakeshima}, \citenamefont {Taira}, \citenamefont {Hinatsu}, \citenamefont {Tobo}, \citenamefont {Ohoyama},\ and\ \citenamefont {Yamaguchi}}]{03_wakeshima}%
  \BibitemOpen
  \bibfield  {author} {\bibinfo {author} {\bibfnamefont {M.}~\bibnamefont {Wakeshima}}, \bibinfo {author} {\bibfnamefont {N.}~\bibnamefont {Taira}}, \bibinfo {author} {\bibfnamefont {Y.}~\bibnamefont {Hinatsu}}, \bibinfo {author} {\bibfnamefont {A.}~\bibnamefont {Tobo}}, \bibinfo {author} {\bibfnamefont {K.}~\bibnamefont {Ohoyama}},\ and\ \bibinfo {author} {\bibfnamefont {Y.}~\bibnamefont {Yamaguchi}},\ }\bibfield  {title} {\bibinfo {title} {Specific heat and neutron diffraction study on quaternary sulfides {BaNd$_2$CoS$_5$} and {BaNd$_2$ZnS$_5$}},\ }\href@noop {} {\bibfield  {journal} {\bibinfo  {journal} {Journal of Solid State Chemistry}\ }\textbf {\bibinfo {volume} {174}},\ \bibinfo {pages} {159} (\bibinfo {year} {2003})}\BibitemShut {NoStop}%
\bibitem [{\citenamefont {Ozawa}\ \emph {et~al.}(2008)\citenamefont {Ozawa}, \citenamefont {Taniguchi}, \citenamefont {Kawaji}, \citenamefont {Mizusaki}, \citenamefont {Nagata}, \citenamefont {Noro}, \citenamefont {Samata}, \citenamefont {Mitamura},\ and\ \citenamefont {Takayanagi}}]{08_ozawa}%
  \BibitemOpen
  \bibfield  {author} {\bibinfo {author} {\bibfnamefont {T.}~\bibnamefont {Ozawa}}, \bibinfo {author} {\bibfnamefont {T.}~\bibnamefont {Taniguchi}}, \bibinfo {author} {\bibfnamefont {Y.}~\bibnamefont {Kawaji}}, \bibinfo {author} {\bibfnamefont {S.}~\bibnamefont {Mizusaki}}, \bibinfo {author} {\bibfnamefont {Y.}~\bibnamefont {Nagata}}, \bibinfo {author} {\bibfnamefont {Y.}~\bibnamefont {Noro}}, \bibinfo {author} {\bibfnamefont {H.}~\bibnamefont {Samata}}, \bibinfo {author} {\bibfnamefont {H.}~\bibnamefont {Mitamura}},\ and\ \bibinfo {author} {\bibfnamefont {S.}~\bibnamefont {Takayanagi}},\ }\bibfield  {title} {\bibinfo {title} {Magnetization and specific heat measurement of the shastry–sutherland lattice compounds: {Ln$_2$BaPdO$_5$} {(Ln=La, Pr, Nd, Sm, Eu, Gd, Dy, Ho)}},\ }\href@noop {} {\bibfield  {journal} {\bibinfo  {journal} {Journal of Alloys and Compounds}\ }\textbf {\bibinfo {volume} {448}},\ \bibinfo {pages} {96} (\bibinfo {year} {2008})}\BibitemShut {NoStop}%
\bibitem [{\citenamefont {Ishii}\ \emph {et~al.}(2020)\citenamefont {Ishii}, \citenamefont {Chen}, \citenamefont {Yoshida}, \citenamefont {Oda}, \citenamefont {Christianson},\ and\ \citenamefont {Yamaura}}]{20_ishii}%
  \BibitemOpen
  \bibfield  {author} {\bibinfo {author} {\bibfnamefont {Y.}~\bibnamefont {Ishii}}, \bibinfo {author} {\bibfnamefont {J.}~\bibnamefont {Chen}}, \bibinfo {author} {\bibfnamefont {H.~K.}\ \bibnamefont {Yoshida}}, \bibinfo {author} {\bibfnamefont {M.}~\bibnamefont {Oda}}, \bibinfo {author} {\bibfnamefont {A.~D.}\ \bibnamefont {Christianson}},\ and\ \bibinfo {author} {\bibfnamefont {K.}~\bibnamefont {Yamaura}},\ }\bibfield  {title} {\bibinfo {title} {High-pressure synthesis, crystal structure, and magnetic properties of the shastry-sutherland-lattice oxides {BaLn$_2$ZnO$_5$} {(Ln = Pr, Sm, Eu)}},\ }\href@noop {} {\bibfield  {journal} {\bibinfo  {journal} {Journal of Solid State Chemistry}\ }\textbf {\bibinfo {volume} {289}},\ \bibinfo {pages} {121489} (\bibinfo {year} {2020})}\BibitemShut {NoStop}%
\bibitem [{\citenamefont {Ishii}\ \emph {et~al.}(2021)\citenamefont {Ishii}, \citenamefont {Sala}, \citenamefont {Stone}, \citenamefont {Garlea}, \citenamefont {Calder}, \citenamefont {Chen}, \citenamefont {Yoshida}, \citenamefont {Fukuoka}, \citenamefont {Yan}, \citenamefont {dela Cruz}, \citenamefont {Du}, \citenamefont {Parker}, \citenamefont {Zhang}, \citenamefont {Batista}, \citenamefont {Yamaura},\ and\ \citenamefont {Christianson}}]{21_ishii}%
  \BibitemOpen
  \bibfield  {author} {\bibinfo {author} {\bibfnamefont {Y.}~\bibnamefont {Ishii}}, \bibinfo {author} {\bibfnamefont {G.}~\bibnamefont {Sala}}, \bibinfo {author} {\bibfnamefont {M.~B.}\ \bibnamefont {Stone}}, \bibinfo {author} {\bibfnamefont {V.~O.}\ \bibnamefont {Garlea}}, \bibinfo {author} {\bibfnamefont {S.}~\bibnamefont {Calder}}, \bibinfo {author} {\bibfnamefont {J.}~\bibnamefont {Chen}}, \bibinfo {author} {\bibfnamefont {H.~K.}\ \bibnamefont {Yoshida}}, \bibinfo {author} {\bibfnamefont {S.}~\bibnamefont {Fukuoka}}, \bibinfo {author} {\bibfnamefont {J.}~\bibnamefont {Yan}}, \bibinfo {author} {\bibfnamefont {C.}~\bibnamefont {dela Cruz}}, \bibinfo {author} {\bibfnamefont {M.-H.}\ \bibnamefont {Du}}, \bibinfo {author} {\bibfnamefont {D.~S.}\ \bibnamefont {Parker}}, \bibinfo {author} {\bibfnamefont {H.}~\bibnamefont {Zhang}}, \bibinfo {author} {\bibfnamefont {C.~D.}\ \bibnamefont {Batista}}, \bibinfo {author} {\bibfnamefont {K.}~\bibnamefont {Yamaura}},\ and\ \bibinfo {author} {\bibfnamefont {A.~D.}\
  \bibnamefont {Christianson}},\ }\bibfield  {title} {\bibinfo {title} {Magnetic properties of the shastry-sutherland lattice material {BaNd$_2$ZnS$_5$}},\ }\href@noop {} {\bibfield  {journal} {\bibinfo  {journal} {Phys. Rev. Mater.}\ }\textbf {\bibinfo {volume} {5}},\ \bibinfo {pages} {064418} (\bibinfo {year} {2021})}\BibitemShut {NoStop}%
\bibitem [{\citenamefont {Billingsley}\ \emph {et~al.}(2022)\citenamefont {Billingsley}, \citenamefont {Marshall}, \citenamefont {Shu}, \citenamefont {Cao},\ and\ \citenamefont {Kong}}]{22_billingsley}%
  \BibitemOpen
  \bibfield  {author} {\bibinfo {author} {\bibfnamefont {B.~R.}\ \bibnamefont {Billingsley}}, \bibinfo {author} {\bibfnamefont {M.}~\bibnamefont {Marshall}}, \bibinfo {author} {\bibfnamefont {Z.}~\bibnamefont {Shu}}, \bibinfo {author} {\bibfnamefont {H.}~\bibnamefont {Cao}},\ and\ \bibinfo {author} {\bibfnamefont {T.}~\bibnamefont {Kong}},\ }\bibfield  {title} {\bibinfo {title} {Single crystal synthesis and magnetic properties of a shastry-sutherland lattice compound ${\mathrm{band}}_{2}{\mathrm{zns}}_{5}$},\ }\href@noop {} {\bibfield  {journal} {\bibinfo  {journal} {Phys. Rev. Mater.}\ }\textbf {\bibinfo {volume} {6}},\ \bibinfo {pages} {104403} (\bibinfo {year} {2022})}\BibitemShut {NoStop}%
\bibitem [{\citenamefont {Marshall}\ \emph {et~al.}(2023)\citenamefont {Marshall}, \citenamefont {Billingsley}, \citenamefont {Bai}, \citenamefont {Ma}, \citenamefont {Kong},\ and\ \citenamefont {Cao}}]{23_marshall}%
  \BibitemOpen
  \bibfield  {author} {\bibinfo {author} {\bibfnamefont {M.}~\bibnamefont {Marshall}}, \bibinfo {author} {\bibfnamefont {B.~R.}\ \bibnamefont {Billingsley}}, \bibinfo {author} {\bibfnamefont {X.}~\bibnamefont {Bai}}, \bibinfo {author} {\bibfnamefont {Q.}~\bibnamefont {Ma}}, \bibinfo {author} {\bibfnamefont {T.}~\bibnamefont {Kong}},\ and\ \bibinfo {author} {\bibfnamefont {H.}~\bibnamefont {Cao}},\ }\bibfield  {title} {\bibinfo {title} {Field-induced partial disorder in a shastry-sutherland lattice},\ }\href@noop {} {\bibfield  {journal} {\bibinfo  {journal} {Nature Communications}\ }\textbf {\bibinfo {volume} {14}} (\bibinfo {year} {2023})}\BibitemShut {NoStop}%
\bibitem [{\citenamefont {Pasco}\ \emph {et~al.}(2023)\citenamefont {Pasco}, \citenamefont {Rai}, \citenamefont {Frontzek}, \citenamefont {Sala}, \citenamefont {Stone}, \citenamefont {Chakoumakos}, \citenamefont {Garlea}, \citenamefont {Christianson},\ and\ \citenamefont {May}}]{23_pasco}%
  \BibitemOpen
  \bibfield  {author} {\bibinfo {author} {\bibfnamefont {C.~M.}\ \bibnamefont {Pasco}}, \bibinfo {author} {\bibfnamefont {B.~K.}\ \bibnamefont {Rai}}, \bibinfo {author} {\bibfnamefont {M.}~\bibnamefont {Frontzek}}, \bibinfo {author} {\bibfnamefont {G.}~\bibnamefont {Sala}}, \bibinfo {author} {\bibfnamefont {M.~B.}\ \bibnamefont {Stone}}, \bibinfo {author} {\bibfnamefont {B.~C.}\ \bibnamefont {Chakoumakos}}, \bibinfo {author} {\bibfnamefont {V.~O.}\ \bibnamefont {Garlea}}, \bibinfo {author} {\bibfnamefont {A.~D.}\ \bibnamefont {Christianson}},\ and\ \bibinfo {author} {\bibfnamefont {A.~F.}\ \bibnamefont {May}},\ }\bibfield  {title} {\bibinfo {title} {Anisotropic magnetism of the shastry-sutherland lattice material {BaNd$_2$PtO$_5$}},\ }\href@noop {} {\bibfield  {journal} {\bibinfo  {journal} {Phys. Rev. Mater.}\ }\textbf {\bibinfo {volume} {7}},\ \bibinfo {pages} {074407} (\bibinfo {year} {2023})}\BibitemShut {NoStop}%
\bibitem [{\citenamefont {Ashtar}\ \emph {et~al.}(2021)\citenamefont {Ashtar}, \citenamefont {Bai}, \citenamefont {Xu}, \citenamefont {Wan}, \citenamefont {Wei}, \citenamefont {Liu}, \citenamefont {Marwat},\ and\ \citenamefont {Tian}}]{21_ashtar}%
  \BibitemOpen
  \bibfield  {author} {\bibinfo {author} {\bibfnamefont {M.}~\bibnamefont {Ashtar}}, \bibinfo {author} {\bibfnamefont {Y.}~\bibnamefont {Bai}}, \bibinfo {author} {\bibfnamefont {L.}~\bibnamefont {Xu}}, \bibinfo {author} {\bibfnamefont {Z.}~\bibnamefont {Wan}}, \bibinfo {author} {\bibfnamefont {Z.}~\bibnamefont {Wei}}, \bibinfo {author} {\bibfnamefont {Y.}~\bibnamefont {Liu}}, \bibinfo {author} {\bibfnamefont {M.~A.}\ \bibnamefont {Marwat}},\ and\ \bibinfo {author} {\bibfnamefont {Z.}~\bibnamefont {Tian}},\ }\bibfield  {title} {\bibinfo {title} {Structure and magnetic properties of melilite-type compounds {RE$_2$Be$_2$GeO7} {(RE = Pr, Nd, Gd–Yb)} with rare-earth ions on shastry–sutherland lattice},\ }\href@noop {} {\bibfield  {journal} {\bibinfo  {journal} {Inorganic Chemistry}\ }\textbf {\bibinfo {volume} {60}},\ \bibinfo {pages} {3626} (\bibinfo {year} {2021})},\ \bibinfo {note} {doi: 10.1021/acs.inorgchem.0c03131}\BibitemShut {NoStop}%
\bibitem [{\citenamefont {Brassington}\ \emph {et~al.}(2024)\citenamefont {Brassington}, \citenamefont {Huang}, \citenamefont {Aczel},\ and\ \citenamefont {Zhou}}]{Brassington_2024}%
  \BibitemOpen
  \bibfield  {author} {\bibinfo {author} {\bibfnamefont {A.}~\bibnamefont {Brassington}}, \bibinfo {author} {\bibfnamefont {Q.}~\bibnamefont {Huang}}, \bibinfo {author} {\bibfnamefont {A.~A.}\ \bibnamefont {Aczel}},\ and\ \bibinfo {author} {\bibfnamefont {H.~D.}\ \bibnamefont {Zhou}},\ }\bibfield  {title} {\bibinfo {title} {Synthesis and magnetic properties of the shastry-sutherland family {R$_2$Be$_2$SiO$_7$} {(R=Nd,Sm,Gd-Yb)}},\ }\href@noop {} {\bibfield  {journal} {\bibinfo  {journal} {Phys. Rev. Mater.}\ }\textbf {\bibinfo {volume} {8}},\ \bibinfo {pages} {014005} (\bibinfo {year} {2024})}\BibitemShut {NoStop}%
\bibitem [{\citenamefont {Liu}\ \emph {et~al.}(2024)\citenamefont {Liu}, \citenamefont {Song}, \citenamefont {Cao}, \citenamefont {Ge}, \citenamefont {Bu}, \citenamefont {Zhou}, \citenamefont {Qin}, \citenamefont {Zeng}, \citenamefont {Li}, \citenamefont {Ling}, \citenamefont {Tong}, \citenamefont {Sheng}, \citenamefont {Yang}, \citenamefont {Wu}, \citenamefont {Guo},\ and\ \citenamefont {Tian}}]{24_liu}%
  \BibitemOpen
  \bibfield  {author} {\bibinfo {author} {\bibfnamefont {A.}~\bibnamefont {Liu}}, \bibinfo {author} {\bibfnamefont {F.}~\bibnamefont {Song}}, \bibinfo {author} {\bibfnamefont {Y.}~\bibnamefont {Cao}}, \bibinfo {author} {\bibfnamefont {H.}~\bibnamefont {Ge}}, \bibinfo {author} {\bibfnamefont {H.}~\bibnamefont {Bu}}, \bibinfo {author} {\bibfnamefont {J.}~\bibnamefont {Zhou}}, \bibinfo {author} {\bibfnamefont {Y.}~\bibnamefont {Qin}}, \bibinfo {author} {\bibfnamefont {Q.}~\bibnamefont {Zeng}}, \bibinfo {author} {\bibfnamefont {J.}~\bibnamefont {Li}}, \bibinfo {author} {\bibfnamefont {L.}~\bibnamefont {Ling}}, \bibinfo {author} {\bibfnamefont {W.}~\bibnamefont {Tong}}, \bibinfo {author} {\bibfnamefont {J.}~\bibnamefont {Sheng}}, \bibinfo {author} {\bibfnamefont {M.}~\bibnamefont {Yang}}, \bibinfo {author} {\bibfnamefont {L.}~\bibnamefont {Wu}}, \bibinfo {author} {\bibfnamefont {H.}~\bibnamefont {Guo}},\ and\ \bibinfo {author} {\bibfnamefont {Z.}~\bibnamefont {Tian}},\ }\bibfield  {title} {\bibinfo {title}
  {Distinct magnetic ground states in shastry-sutherland lattice materials: $\mathrm{P}{\mathrm{r}}_{2}\mathrm{B}{\mathrm{e}}_{2}\mathrm{Ge}{\mathrm{o}}_{7}$ versus $\mathrm{N}{\mathrm{d}}_{2}\mathrm{B}{\mathrm{e}}_{2}\mathrm{Ge}{\mathrm{o}}_{7}$},\ }\href {https://link.aps.org/doi/10.1103/PhysRevB.109.184413} {\bibfield  {journal} {\bibinfo  {journal} {Phys. Rev. B}\ }\textbf {\bibinfo {volume} {109}},\ \bibinfo {pages} {184413} (\bibinfo {year} {2024})}\BibitemShut {NoStop}%
\bibitem [{\citenamefont {Rodríguez-Carvajal}(1993)}]{93_rodriguez}%
  \BibitemOpen
  \bibfield  {author} {\bibinfo {author} {\bibfnamefont {J.}~\bibnamefont {Rodríguez-Carvajal}},\ }\bibfield  {title} {\bibinfo {title} {Recent advances in magnetic structure determination by neutron powder diffraction},\ }\href@noop {} {\bibfield  {journal} {\bibinfo  {journal} {Physica B: Condensed Matter}\ }\textbf {\bibinfo {volume} {192}},\ \bibinfo {pages} {55} (\bibinfo {year} {1993})}\BibitemShut {NoStop}%
\bibitem [{\citenamefont {Kuz'micheva}\ \emph {et~al.}(2002)\citenamefont {Kuz'micheva}, \citenamefont {Rybakov}, \citenamefont {Kutovoi}, \citenamefont {Panyutin}, \citenamefont {Oleinik},\ and\ \citenamefont {Plashkarev}}]{Kuz'micheva2002}%
  \BibitemOpen
  \bibfield  {author} {\bibinfo {author} {\bibfnamefont {G.~M.}\ \bibnamefont {Kuz'micheva}}, \bibinfo {author} {\bibfnamefont {V.~B.}\ \bibnamefont {Rybakov}}, \bibinfo {author} {\bibfnamefont {S.~A.}\ \bibnamefont {Kutovoi}}, \bibinfo {author} {\bibfnamefont {V.~L.}\ \bibnamefont {Panyutin}}, \bibinfo {author} {\bibfnamefont {A.~Y.}\ \bibnamefont {Oleinik}},\ and\ \bibinfo {author} {\bibfnamefont {O.~G.}\ \bibnamefont {Plashkarev}},\ }\bibfield  {title} {\bibinfo {title} {Preparation, structure, and properties of new laser crystals {Y$_2$SiBe$_2$O$_7$}and y$_2$al(beb)o$_7$},\ }\href@noop {} {\bibfield  {journal} {\bibinfo  {journal} {Inorganic Materials}\ }\textbf {\bibinfo {volume} {38}},\ \bibinfo {pages} {60} (\bibinfo {year} {2002})}\BibitemShut {NoStop}%
\bibitem [{\citenamefont {Calder}\ \emph {et~al.}(2018)\citenamefont {Calder}, \citenamefont {An}, \citenamefont {Boehler}, \citenamefont {Dela~Cruz}, \citenamefont {Frontzek}, \citenamefont {Guthrie}, \citenamefont {Haberl}, \citenamefont {Huq}, \citenamefont {Kimber}, \citenamefont {Liu}, \citenamefont {Molaison}, \citenamefont {Neuefeind}, \citenamefont {Page}, \citenamefont {dos Santos}, \citenamefont {Taddei}, \citenamefont {Tulk},\ and\ \citenamefont {Tucker}}]{18_calder}%
  \BibitemOpen
  \bibfield  {author} {\bibinfo {author} {\bibfnamefont {S.}~\bibnamefont {Calder}}, \bibinfo {author} {\bibfnamefont {K.}~\bibnamefont {An}}, \bibinfo {author} {\bibfnamefont {R.}~\bibnamefont {Boehler}}, \bibinfo {author} {\bibfnamefont {C.~R.}\ \bibnamefont {Dela~Cruz}}, \bibinfo {author} {\bibfnamefont {M.~D.}\ \bibnamefont {Frontzek}}, \bibinfo {author} {\bibfnamefont {M.}~\bibnamefont {Guthrie}}, \bibinfo {author} {\bibfnamefont {B.}~\bibnamefont {Haberl}}, \bibinfo {author} {\bibfnamefont {A.}~\bibnamefont {Huq}}, \bibinfo {author} {\bibfnamefont {S.~A.~J.}\ \bibnamefont {Kimber}}, \bibinfo {author} {\bibfnamefont {J.}~\bibnamefont {Liu}}, \bibinfo {author} {\bibfnamefont {J.~J.}\ \bibnamefont {Molaison}}, \bibinfo {author} {\bibfnamefont {J.}~\bibnamefont {Neuefeind}}, \bibinfo {author} {\bibfnamefont {K.}~\bibnamefont {Page}}, \bibinfo {author} {\bibfnamefont {A.~M.}\ \bibnamefont {dos Santos}}, \bibinfo {author} {\bibfnamefont {K.~M.}\ \bibnamefont {Taddei}}, \bibinfo {author} {\bibfnamefont
  {C.}~\bibnamefont {Tulk}},\ and\ \bibinfo {author} {\bibfnamefont {M.~G.}\ \bibnamefont {Tucker}},\ }\bibfield  {title} {\bibinfo {title} {{A suite-level review of the neutron powder diffraction instruments at Oak Ridge National Laboratory}},\ }\href@noop {} {\bibfield  {journal} {\bibinfo  {journal} {Review of Scientific Instruments}\ }\textbf {\bibinfo {volume} {89}},\ \bibinfo {pages} {092701} (\bibinfo {year} {2018})}\BibitemShut {NoStop}%
\bibitem [{\citenamefont {Granroth}\ \emph {et~al.}(2010)\citenamefont {Granroth}, \citenamefont {Kolesnikov}, \citenamefont {Sherline}, \citenamefont {Clancy}, \citenamefont {Ross}, \citenamefont {Ruff}, \citenamefont {Gaulin},\ and\ \citenamefont {Nagler}}]{10_granroth}%
  \BibitemOpen
  \bibfield  {author} {\bibinfo {author} {\bibfnamefont {G.~E.}\ \bibnamefont {Granroth}}, \bibinfo {author} {\bibfnamefont {A.~I.}\ \bibnamefont {Kolesnikov}}, \bibinfo {author} {\bibfnamefont {T.~E.}\ \bibnamefont {Sherline}}, \bibinfo {author} {\bibfnamefont {J.~P.}\ \bibnamefont {Clancy}}, \bibinfo {author} {\bibfnamefont {K.~A.}\ \bibnamefont {Ross}}, \bibinfo {author} {\bibfnamefont {J.~P.~C.}\ \bibnamefont {Ruff}}, \bibinfo {author} {\bibfnamefont {B.~D.}\ \bibnamefont {Gaulin}},\ and\ \bibinfo {author} {\bibfnamefont {S.~E.}\ \bibnamefont {Nagler}},\ }\bibfield  {title} {\bibinfo {title} {Sequoia: A newly operating chopper spectrometer at the sns},\ }\href@noop {} {\bibfield  {journal} {\bibinfo  {journal} {Journal of Physics: Conference Series}\ }\textbf {\bibinfo {volume} {251}},\ \bibinfo {pages} {012058} (\bibinfo {year} {2010})}\BibitemShut {NoStop}%
\bibitem [{\citenamefont {Cao}\ \emph {et~al.}(2019)\citenamefont {Cao}, \citenamefont {Chakoumakos}, \citenamefont {Andrews}, \citenamefont {Wu}, \citenamefont {Riedel}, \citenamefont {Hodges}, \citenamefont {Zhou}, \citenamefont {Gregory}, \citenamefont {Haberl}, \citenamefont {Molaison},\ and\ \citenamefont {Lynn}}]{19_cao}%
  \BibitemOpen
  \bibfield  {author} {\bibinfo {author} {\bibfnamefont {H.}~\bibnamefont {Cao}}, \bibinfo {author} {\bibfnamefont {B.~C.}\ \bibnamefont {Chakoumakos}}, \bibinfo {author} {\bibfnamefont {K.~M.}\ \bibnamefont {Andrews}}, \bibinfo {author} {\bibfnamefont {Y.}~\bibnamefont {Wu}}, \bibinfo {author} {\bibfnamefont {R.~A.}\ \bibnamefont {Riedel}}, \bibinfo {author} {\bibfnamefont {J.}~\bibnamefont {Hodges}}, \bibinfo {author} {\bibfnamefont {W.}~\bibnamefont {Zhou}}, \bibinfo {author} {\bibfnamefont {R.}~\bibnamefont {Gregory}}, \bibinfo {author} {\bibfnamefont {B.}~\bibnamefont {Haberl}}, \bibinfo {author} {\bibfnamefont {J.}~\bibnamefont {Molaison}},\ and\ \bibinfo {author} {\bibfnamefont {G.~W.}\ \bibnamefont {Lynn}},\ }\bibfield  {title} {\bibinfo {title} {Demand, a dimensional extreme magnetic neutron diffractometer at the high flux isotope reactor},\ }\href@noop {} {\bibfield  {journal} {\bibinfo  {journal} {Crystals}\ }\textbf {\bibinfo {volume} {9}} (\bibinfo {year} {2019})}\BibitemShut {NoStop}%
\bibitem [{\citenamefont {Ochiai}\ \emph {et~al.}(2011)\citenamefont {Ochiai}, \citenamefont {Matsuda}, \citenamefont {Ikeda}, \citenamefont {Shimizu}, \citenamefont {Toyoshima}, \citenamefont {Aoki},\ and\ \citenamefont {Katoh}}]{Ochiai_2011}%
  \BibitemOpen
  \bibfield  {author} {\bibinfo {author} {\bibfnamefont {A.}~\bibnamefont {Ochiai}}, \bibinfo {author} {\bibfnamefont {S.}~\bibnamefont {Matsuda}}, \bibinfo {author} {\bibfnamefont {Y.}~\bibnamefont {Ikeda}}, \bibinfo {author} {\bibfnamefont {Y.}~\bibnamefont {Shimizu}}, \bibinfo {author} {\bibfnamefont {S.}~\bibnamefont {Toyoshima}}, \bibinfo {author} {\bibfnamefont {H.}~\bibnamefont {Aoki}},\ and\ \bibinfo {author} {\bibfnamefont {K.}~\bibnamefont {Katoh}},\ }\bibfield  {title} {\bibinfo {title} {Field-induced partially disordered state in {Yb$_2$Pt$_2$Pb}},\ }\href@noop {} {\bibfield  {journal} {\bibinfo  {journal} {Journal of the Physical Society of Japan}\ }\textbf {\bibinfo {volume} {80}},\ \bibinfo {pages} {123705} (\bibinfo {year} {2011})}\BibitemShut {NoStop}%
\bibitem [{\citenamefont {Gannon}\ \emph {et~al.}(2019{\natexlab{b}})\citenamefont {Gannon}, \citenamefont {Zaliznyak}, \citenamefont {Wu}, \citenamefont {Feiguin}, \citenamefont {Tsvelik}, \citenamefont {Demmel}, \citenamefont {Qiu}, \citenamefont {Copley}, \citenamefont {Kim},\ and\ \citenamefont {Aronson}}]{Gannon_2019}%
  \BibitemOpen
  \bibfield  {author} {\bibinfo {author} {\bibfnamefont {W.~J.}\ \bibnamefont {Gannon}}, \bibinfo {author} {\bibfnamefont {I.~A.}\ \bibnamefont {Zaliznyak}}, \bibinfo {author} {\bibfnamefont {L.~S.}\ \bibnamefont {Wu}}, \bibinfo {author} {\bibfnamefont {A.~E.}\ \bibnamefont {Feiguin}}, \bibinfo {author} {\bibfnamefont {A.~M.}\ \bibnamefont {Tsvelik}}, \bibinfo {author} {\bibfnamefont {F.}~\bibnamefont {Demmel}}, \bibinfo {author} {\bibfnamefont {Y.}~\bibnamefont {Qiu}}, \bibinfo {author} {\bibfnamefont {J.~R.~D.}\ \bibnamefont {Copley}}, \bibinfo {author} {\bibfnamefont {M.~S.}\ \bibnamefont {Kim}},\ and\ \bibinfo {author} {\bibfnamefont {M.~C.}\ \bibnamefont {Aronson}},\ }\bibfield  {title} {\bibinfo {title} {Spinon confinement and a sharp longitudinal mode in {Yb$_2$Pt$_2$Pb} in magnetic fields},\ }\href {http://dx.doi.org/10.1038/s41467-019-08715-y} {\bibfield  {journal} {\bibinfo  {journal} {Nature Communications}\ }\textbf {\bibinfo {volume} {10}} (\bibinfo {year} {2019}{\natexlab{b}})}\BibitemShut
  {NoStop}%
\bibitem [{\citenamefont {Hutchings}(1964)}]{Hutchings_1964}%
  \BibitemOpen
  \bibfield  {author} {\bibinfo {author} {\bibfnamefont {M.}~\bibnamefont {Hutchings}},\ }\bibfield  {title} {\bibinfo {title} {Point-charge calculations of energy levels of magnetic ions in crystalline electric fields}\ }(\bibinfo  {publisher} {Academic Press},\ \bibinfo {year} {1964})\ pp.\ \bibinfo {pages} {227--273}\BibitemShut {NoStop}%
\bibitem [{\citenamefont {Scheie}\ \emph {et~al.}(2018)\citenamefont {Scheie}, \citenamefont {Sanders}, \citenamefont {Krizan}, \citenamefont {Christianson}, \citenamefont {Garlea}, \citenamefont {Cava},\ and\ \citenamefont {Broholm}}]{Scheie_2018}%
  \BibitemOpen
  \bibfield  {author} {\bibinfo {author} {\bibfnamefont {A.}~\bibnamefont {Scheie}}, \bibinfo {author} {\bibfnamefont {M.}~\bibnamefont {Sanders}}, \bibinfo {author} {\bibfnamefont {J.}~\bibnamefont {Krizan}}, \bibinfo {author} {\bibfnamefont {A.~D.}\ \bibnamefont {Christianson}}, \bibinfo {author} {\bibfnamefont {V.~O.}\ \bibnamefont {Garlea}}, \bibinfo {author} {\bibfnamefont {R.~J.}\ \bibnamefont {Cava}},\ and\ \bibinfo {author} {\bibfnamefont {C.}~\bibnamefont {Broholm}},\ }\bibfield  {title} {\bibinfo {title} {Crystal field levels and magnetic anisotropy in the kagome compounds {${\mathrm{Nd}}_{3}{\mathrm{Sb}}_{3}{\mathrm{Mg}}_{2}{\mathrm{O}}_{14}$, ${\mathrm{Nd}}_{3}{\mathrm{Sb}}_{3}{\mathrm{Zn}}_{2}{\mathrm{O}}_{14}$, and ${\mathrm{Pr}}_{3}{\mathrm{Sb}}_{3}{\mathrm{Mg}}_{2}{\mathrm{O}}_{14}$}},\ }\href@noop {} {\bibfield  {journal} {\bibinfo  {journal} {Phys. Rev. B}\ }\textbf {\bibinfo {volume} {98}},\ \bibinfo {pages} {134401} (\bibinfo {year} {2018})}\BibitemShut {NoStop}%
\bibitem [{\citenamefont {Scheie}(2021)}]{21_scheie}%
  \BibitemOpen
  \bibfield  {author} {\bibinfo {author} {\bibfnamefont {A.}~\bibnamefont {Scheie}},\ }\bibfield  {title} {\bibinfo {title} {{{\it PyCrystalField}: software for calculation, analysis and fitting of crystal electric field Hamiltonians}},\ }\href@noop {} {\bibfield  {journal} {\bibinfo  {journal} {Journal of Applied Crystallography}\ }\textbf {\bibinfo {volume} {54}},\ \bibinfo {pages} {356} (\bibinfo {year} {2021})}\BibitemShut {NoStop}%
\bibitem [{\citenamefont {Stevens}(1952)}]{Stevens}%
  \BibitemOpen
  \bibfield  {author} {\bibinfo {author} {\bibfnamefont {K.~W.~H.}\ \bibnamefont {Stevens}},\ }\href {https://dx.doi.org/10.1088/0370-1298/65/3/308} {\ \textbf {\bibinfo {volume} {65}},\ \bibinfo {pages} {209} (\bibinfo {year} {1952})}\BibitemShut {NoStop}%
\bibitem [{\citenamefont {Jensen}\ and\ \citenamefont {Mackintosh}(1991)}]{Jensen_1991}%
  \BibitemOpen
  \bibfield  {author} {\bibinfo {author} {\bibfnamefont {J.}~\bibnamefont {Jensen}}\ and\ \bibinfo {author} {\bibfnamefont {A.~R.}\ \bibnamefont {Mackintosh}},\ }\href@noop {} {\emph {\bibinfo {title} {Rare earth magnetism: Structures and excitations}}}\ (\bibinfo  {publisher} {Clarendon Press},\ \bibinfo {year} {1991})\BibitemShut {NoStop}%
\bibitem [{\citenamefont {Ma}\ \emph {et~al.}(2023)\citenamefont {Ma}, \citenamefont {Bai}, \citenamefont {Feng}, \citenamefont {Zhang},\ and\ \citenamefont {Cao}}]{MaCEF}%
  \BibitemOpen
  \bibfield  {author} {\bibinfo {author} {\bibfnamefont {Q.}~\bibnamefont {Ma}}, \bibinfo {author} {\bibfnamefont {X.}~\bibnamefont {Bai}}, \bibinfo {author} {\bibfnamefont {E.}~\bibnamefont {Feng}}, \bibinfo {author} {\bibfnamefont {G.}~\bibnamefont {Zhang}},\ and\ \bibinfo {author} {\bibfnamefont {H.}~\bibnamefont {Cao}},\ }\bibfield  {title} {\bibinfo {title} {{{\it CrysFieldExplorer}: rapid optimization of the crystal field Hamiltonian}},\ }\href@noop {} {\bibfield  {journal} {\bibinfo  {journal} {Journal of Applied Crystallography}\ }\textbf {\bibinfo {volume} {56}},\ \bibinfo {pages} {1229} (\bibinfo {year} {2023})}\BibitemShut {NoStop}%
\bibitem [{08_(2008)}]{08_gruber}%
  \BibitemOpen
  \bibfield  {title} {\bibinfo {title} {Spectroscopic and magnetic susceptibility analyses of the 7fj and 5d4 energy levels of {Tb3+(4f8)} in {TbAlO3}},\ }\href@noop {} {\bibfield  {journal} {\bibinfo  {journal} {Journal of Luminescence}\ }\textbf {\bibinfo {volume} {128}},\ \bibinfo {pages} {1271} (\bibinfo {year} {2008})}\BibitemShut {NoStop}%
\bibitem [{\citenamefont {Brassington}(2024)}]{brassington_unpublished}%
  \BibitemOpen
  \bibfield  {author} {\bibinfo {author} {\bibfnamefont {A.}~\bibnamefont {Brassington}}} (\bibinfo {year} {2024}),\ \bibinfo {note} {unpublished}\BibitemShut {NoStop}%
\bibitem [{\citenamefont {Wills}(2000)}]{WILLS_2006}%
  \BibitemOpen
  \bibfield  {author} {\bibinfo {author} {\bibfnamefont {A.}~\bibnamefont {Wills}},\ }\bibfield  {title} {\bibinfo {title} {A new protocol for the determination of magnetic structures using simulated annealing and representational analysis (sarah)},\ }\href@noop {} {\bibfield  {journal} {\bibinfo  {journal} {Physica B: Condensed Matter}\ }\textbf {\bibinfo {volume} {276-278}},\ \bibinfo {pages} {680} (\bibinfo {year} {2000})}\BibitemShut {NoStop}%
\bibitem [{\citenamefont {Kovalev}(1993)}]{93_kovalev}%
  \BibitemOpen
  \bibfield  {author} {\bibinfo {author} {\bibfnamefont {O.}~\bibnamefont {Kovalev}},\ }\href@noop {} {\emph {\bibinfo {title} {Representations of the Crystallographic Space Groups, 2nd ed.}}}\ (\bibinfo  {publisher} {Gordon and Breach},\ \bibinfo {address} {Switzerland},\ \bibinfo {year} {1993})\BibitemShut {NoStop}%
\end{thebibliography}%

\end{document}